\documentclass[lettersize,journal]{IEEEtran}
\usepackage{amsmath,amsfonts}
\usepackage{algorithmic}
\usepackage{array}
\usepackage[caption=false,font=normalsize,labelfont=sf,textfont=sf]{subfig}
\usepackage{textcomp}
\usepackage{stfloats}
\usepackage{url}
\usepackage{verbatim}
\usepackage{graphicx}
\usepackage{xcolor}
\usepackage{amsmath} 
\usepackage{amsmath, amssymb, amsthm}
\usepackage{amsthm}
\usepackage{comment}
\usepackage{stfloats}
\usepackage{balance}
\usepackage{epstopdf}
\usepackage{subcaption}
\usepackage{caption}    
\usepackage{cite}

\belowdisplayskip=\abovedisplayskip
\def\BibTeX{{\rm B\kern-.05em{\sc i\kern-.025em b}\kern-.08em
    T\kern-.1667em\lower.7ex\hbox{E}\kern-.125emX}}

\begin{document}
\title{Analysis of Fluid Antenna Systems with Continuous Positioning and Spatial Correlation}
\author{IEEE Publication Technology Department
\thanks{Manuscript created October, 2020; This work was developed by the IEEE Publication Technology Department. This work is distributed under the \LaTeX \ Project Public License (LPPL) ( http://www.latex-project.org/ ) version 1.3. A copy of the LPPL, version 1.3, is included in the base \LaTeX \ documentation of all distributions of \LaTeX \ released 2003/12/01 or later. The opinions expressed here are entirely that of the author. No warranty is expressed or implied. User assumes all risk.}}

\author{Gayani~Siriwardana,~\IEEEmembership{Student Member,~IEEE,}       
        Peter~J.~Smith,~\IEEEmembership{Fellow,~IEEE,}
        Himal~A.~Suraweera,~\IEEEmembership{Senior Member,~IEEE,}
        and~Rajitha~Senanayake,~\IEEEmembership{Member,~IEEE}. 
\thanks{G. Siriwardana and R. Senanayake are with the Department of Electrical
and Electronic Engineering, University of Melbourne, Parkville,
VIC. 3010, Australia (e-mail: gayani.siriwardana@student.unimelb.edu.au, rajitha.senanayake@unimelb.edu.au).}
\thanks{P. J. Smith is with the School of Mathematics and Statistics, Victoria University of Wellington, Wellington, New Zealand (e-mail: peter.smith@vuw.ac.nz).}
\thanks{H. A. Suraweera is with the Department of Electrical
and Electronic Engineering, University of Peradeniya, Peradeniya 20400, Sri Lanka (e-mail: himal@eng.pdn.ac.lk).}}

\maketitle

\begin{abstract}
We analyze multi-user fluid antenna systems with continuous positioning over a track of length $L$ under a spatial correlation model, where exact performance distributions become analytically intractable. We develop a level-crossing-rate (LCR) framework that yields asymptotically exact approximations and tight bounds for the cumulative distribution function (cdf) of the optimized metric $S^\star=\sup_{0\le l\le L} S(l)$, where $S(l)$ denotes the performance metric at antenna position $l$. For a single fluid antenna, we characterize the cdfs of signal-to-noise ratio (SNR), signal-to-interference ratio (SIR) and signal-to-interference-plus-noise ratio (SINR) under Rayleigh fading and extend the approach to Ricean desired channels. We further treat two multi-antenna receiver layouts with maximum-ratio combining: (i) a fluid antenna with a fixed antenna and (ii) a two-element moving array, deriving new LCR results for the practically important case where array-element correlation and positional correlation are inherently coupled. The analysis provides actionable insights: high-threshold tail probabilities scale linearly with $L$, we derive the required $L$ to neutralize a co-channel interferer, and we show that about one wavelength of movement can reduce outage by three orders of magnitude. Monte Carlo results validate the accuracy across the considered scenarios and regimes.
\end{abstract}


\section{Introduction}\label{sec:intro}

Advances in multiple-antenna technology, and especially antenna arrays, have been central to the evolution of modern wireless communications \cite{zhang20}. As time and frequency resources become increasingly congested, there is growing emphasis on exploiting the \emph{spatial} dimension to increase spectral efficiency and reliability. Multiple-input multiple-output (MIMO) \cite{Paulraj94} and massive MIMO \cite{Larsson14} deliver substantial gains by leveraging spatial multiplexing and diversity and are now integral to contemporary wireless systems. However, conventional MIMO architectures typically employ \emph{fixed} antenna arrays, which constrains the achievable spatial diversity and array geometry to a predetermined layout within a limited physical aperture. To unlock greater flexibility in the spatial domain---and thereby increase spatial degrees of freedom---reconfigurable antenna concepts are required \cite{WeeKiat_ICST_2024}.

One emerging reconfigurable paradigm with significant potential is the \emph{fluid antenna system} (FAS). A fluid antenna refers to a software-controlled structure composed of fluidic, conductive, or dielectric material that can dynamically reshape and/or reposition its radiating structure \cite{WeeKiat_ICST_2024}. This broad concept encompasses multiple realizations of position-flexible antennas, including pixel-based on--off switching structures \cite{Cetiner_ICM_2004,Zhang_IOJAP_2024}, liquid-based antennas \cite{Huang_IOJAP_2021,Borda-Fortuny_IA_2019}, and mechanically movable antenna systems \cite{Zhu_ICM_2024,Lipeng_2024}. By enabling controlled adaptation of the effective antenna position, aperture, and radiation properties, FAS technology has the potential to improve spectrum utilization, increase spatial diversity within compact form factors, and support adaptive functions such as beamforming and polarization control \cite{WeeKiat_ICST_2024}.

\subsection{Background and Motivation}
A rapidly expanding body of work now exists on FASs across diverse scenarios, including single-/multi-user, uplink/downlink, and single-/multi-antenna settings. Recent surveys and tutorials provide comprehensive state-of-the-art overviews \cite{WeeKiat_ICST_2024,survey}. In the following, we highlight three aspects that are central to this paper: (i) spatial correlation modeling, (ii) continuous positioning, and (iii) mathematical performance analysis of FASs.

\subsubsection{Spatial Correlation}
In many FAS implementations the available antenna positions (i.e., ports) are separated by small distances due to physical size constraints. This issue is particularly pronounced under continuous positioning, where infinitely many positions are available over a finite track. As a result, accurate modeling of spatial correlation is essential for revealing the true diversity and interference-mitigation capabilities of FASs. Most statistical analyses start from correlation models that depend only on spatial separation---with Jakes' model being a common choice \cite{WeeKiat_ICST_2024,Wong20,Wong_ITWC_2021}. However, it is well recognized that a full analytical treatment under such physical correlation structures can be \emph{prohibitively complex} \cite{Ramírez-Espinosa_TWC_2024}. Consequently, a range of tractable approximations to the correlation structure has been developed. Examples include reference-based correlation constructions \cite{Wong_ITWC_2021}, two-stage approximations that reduce key performance expressions (e.g., outage) to a single integral \cite{Khammassi_TWC_2023}, and block-diagonal correlation frameworks \cite{Ramírez-Espinosa_TWC_2024} that build on averaged-parameter approaches \cite{Wong_IEL_2022}. In parallel, copula-based methods have been proposed to preserve marginal distributions while capturing dependence \cite{copula2,copula3}. While these approaches can be effective numerically, their usage is restricted to discrete positioning and extension under continuous positioning remains an open question.

\subsubsection{Continuous Positioning}
Most existing work considers fluid antennas that select among a finite set of $n$ discrete ports \cite{Wong_ITWC_2021,Wong20,Wong_IWC_2022,Psomas23}. Discrete-port FASs can approach near-maximum performance with relatively few ports \cite{weeKiat_2024TWC}, but continuous positioning is a valuable theoretical benchmark: it provides an upper bound on achievable performance and enables exploration of the full spatial diversity available over a given track length. Importantly, practical mechanisms for future implementations may also support effectively continuous placement within a constrained region \cite{WeeKiat_ICST_2024,Psomas_ICL_2023,statCSI,multi,UL,DL}. Analytical characterization for continuous-position FASs is limited; \cite{Psomas_ICL_2023} derives a lower bound on the signal-to-interference
ratio (SIR), whereas much of the remaining literature focuses on optimization and algorithm design under complex system constraints \cite{statCSI,multi,UL,DL}. This motivates further analytical development for continuous-position models, which we pursue in this work.

\subsubsection{Mathematical Performance Analysis}
When a physical correlation model is used, the correlation between channel responses depends only on their spatial separation and applies consistently across all location pairs along the track. While desirable from a modeling standpoint, this correlation introduces substantial analytical difficulty. Under Rayleigh fading, the joint distribution of channels across $n$ antenna positions becomes an $n$-dimensional multivariate Rayleigh distribution. Performance metrics based on selecting the best position (e.g., maximizing signal-to-noise ratio (SNR)) therefore depend on the maximum of \emph{correlated} fading gains, and computing the corresponding cumulative distribution function (cdf) requires integration over the joint probability density function (pdf). This is tractable only for small $n$: for one port the SNR is exponential; for two ports the bivariate density involves a single Bessel function \cite{bivRayleigh}; and for $n=3$ and $n=4$ the joint pdfs involve increasingly complex infinite-series representations \cite{triRayleigh}. Due to this rapid growth in complexity, much prior work has relied on correlation-structure approximations to enable analytical progress \cite{Khammassi_TWC_2023,Wong_ITWC_2021,Wong20,Ramírez-Espinosa_TWC_2024,copula2,copula3}. Exact analytical characterizations remain rare, although recent results have established the exact diversity order for discrete-port FASs via tractable SNR approximations \cite{Zhao_2025}.

Motivated by the above challenges, this paper develops analytical approximations for key continuous-position FAS performance metrics, including SNR, SIR and signal-to-interference-plus-noise ratio (SINR) under both Rayleigh and Ricean desired channels. The approximations show strong agreement with Monte Carlo simulations across the considered regimes and enable clear insights into two fundamental benefits of FAS technology: improved SNR statistics through spatial variability and interference suppression through spatial adaptation.

\subsection{Contributions}
The core analytical challenges addressed in this paper can be summarized as follows:
\begin{itemize}
    \item Spatial correlation models are physically appealing but render exact analysis of selection-based performance (e.g., the SNR cdf) mathematically intractable for all but very small numbers of ports, motivating tractable correlation approximations \cite{Wong_ITWC_2021,Khammassi_TWC_2023,Ramírez-Espinosa_TWC_2024,copula2,copula3}.
    \item Continuous positioning serves as an upper bound for discrete-port implementations and captures the full spatial diversity of a finite track. While optimization-based studies exist \cite{statCSI,multi,UL,DL}, analytical results are limited, with \cite{Psomas_ICL_2023} being a key exception.
    \item Exact closed-form characterizations of continuous-position FAS performance are scarce, particularly under spatial correlation.
\end{itemize}
To address these challenges, we make the following contributions:
\begin{itemize}
    \item We study three receiver layouts involving single and multiple antennas in a fluid-antenna architecture, under continuous positioning of the movable components with spatial correlation models.
    \item We introduce a novel technique to accurately approximate the cdf of the fluid antenna system, and derive analytically tractable approximations for the cdf of the relevant optimized performance metric (SNR, SIR, or SINR) of each layout. As part of the derivations, new level crossing rate (LCR) results for SNR, SIR, and SINR are also established.
    \item Using the resulting framework, we obtain actionable performance insights. In particular, we show that increasing the movable length by approximately one wavelength can improve outage probability by around three orders of magnitude, derive the movable length required to effectively neutralize a co-channel interferer, and show that the probability of achieving high SNR scales proportionally with the available movement length.
\end{itemize}

The remainder of this paper is organized as follows. Section \ref{sec:single} presents the system model and performance analysis for the single fluid antenna system. Section \ref{multisec} introduces two multi-antenna receiver layouts and their corresponding analyses. Section \ref{numres} provides numerical results for all three layouts, and Section \ref{conc} concludes the paper.

\textbf{\textit{Notation:}} We use boldface upper and lower case letters for matrices and column vectors, respectively. The operator $\mathbb{E}[\cdot]$ denotes the statistical expectation.  $(\cdot)^T$, $|\cdot|$, $(\cdot)^{\mathrm{H}}$ and $(\cdot)^*$ stand for transpose, magnitude, conjugate transpose and complex conjugation respectively. Finally, $\mathcal{CN}$ denotes the circularly-symmetric complex normal distribution and $\mathcal{N}$ denotes the real Gaussian distribution.

\section{Single Fluid Antenna System}\label{sec:single}

\subsection{System Model}
We begin with a point-to-point link in which a single-antenna transmitter communicates with a receiver equipped with a \emph{single fluid antenna}. The fluid antenna can be positioned \emph{continuously} along a one-dimensional track of length $L$ (measured in wavelengths). Let $l\in[0,L]$ denote the FA position. As is common when characterizing the best-achievable performance, we neglect movement delay and associated overheads \cite{Wong_ITWC_2021, Wong_IWC_2022}.

\begin{figure}[t]\centering
  \includegraphics[width=\linewidth]{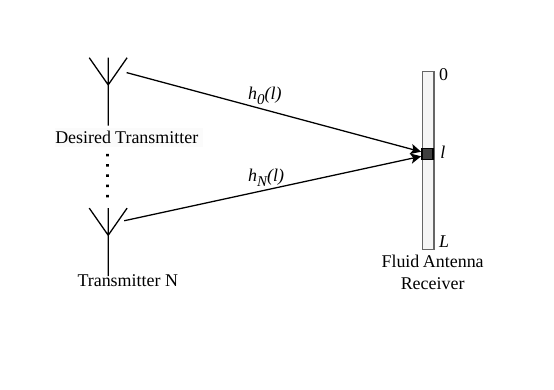}
  \caption{System model of a single fluid antenna receiver with $N$ interferers. Channels from the desired transmitter and interferers are denoted by $h_{0}(l)$ and $h_{i}(l)$, respectively.}
  \captionsetup{justification=centering}
  \label{im1}
\end{figure}

With $N$ co-channel interferers, the complex baseband signal received at position $l$ is modeled as
\begin{equation}
\label{received_signal}
  y(l) = h_{0}(l)x_{0} + \sum_{i=1}^{N} h_{i}(l)x_{i} + n(l),\qquad 0 \leq l \leq L,
\end{equation}
where $h_{0}(l)$ is the desired channel, $h_{i}(l)$ is the channel from the $i$th interferer, $x_0$ and $x_i$ are the corresponding transmitted symbols, and $n(l)$ is additive white Gaussian noise (AWGN). We assume $\{h_i(l)\}_{i=1}^N$ are mutually independent and are independent of $h_0(l)$. The channels are wide-sense stationary along the track, with
$\mathbb{E}[|h_{0}(l)|^2]=\beta_0$ and $\mathbb{E}[|h_{i}(l)|^2]=\beta_i$.
The desired link is modeled as either Rayleigh or Ricean, while each interferer channel is Rayleigh.
The symbols satisfy $\mathbb{E}[|x_0|^2]=E_{x_0}$ and $\mathbb{E}[|x_i|^2]=E_{x_i}$, and $n(l)\sim\mathcal{CN}(0,\sigma^2)$ is independent of all channels and symbols.

\paragraph{Spatial correlation model}
Under a classic isotropic scattering assumption, the correlation between channels from the \emph{same} source (e.g., interferer $i$) at two positions separated by $\tau$ wavelengths is modeled via Jakes' model \cite{Heath_book_2018}
\begin{equation}\label{correlation}
\rho(\tau) 
\triangleq \frac{1}{\beta_i}\,\mathbb{E}\big[h_i(l)h_i^*(l+\tau)\big]
= J_0(2\pi\tau),
\end{equation}
where $J_0(\cdot)$ is the zero-th order Bessel function of the first kind \cite[Eq.\ (8.402)]{Gradshteyn_book_2007}.
It is useful to note that any correlation model satisfying the local expansion
\begin{align}\label{correlation2}
\rho(\tau) = 1 - b\tau^2 + o(\tau^2),\qquad \tau\to 0,\; b>0,
\end{align}
induces a mean-square differentiable (and hence physically reasonable) Rayleigh fading process. The little-$o$ notation is in the standard Bachmann--Landau sense. For the Jakes' model in \eqref{correlation}, the series representation of $J_0(\cdot)$ gives $b=\pi^2$ \cite{Gradshteyn_book_2007}.

\paragraph{Performance metrics and spatial maximization}
When the FA is at position $l$, the instantaneous SNR is
\begin{equation}\label{snr}
\mathrm{SNR}(l) = \frac{E_{x_0}|h_0(l)|^2}{\sigma^2},
\end{equation}
and the corresponding SIR and SINR are
\begin{equation}\label{sir}
\mathrm{SIR}(l) = \frac{E_{x_0}|h_0(l)|^2}{\sum_{i=1}^{N} E_{x_i}|h_i(l)|^2},
\end{equation}
\begin{equation}\label{sinr}
\mathrm{SINR}(l) = \frac{E_{x_0}|h_0(l)|^2}{\sum_{i=1}^{N} E_{x_i}|h_i(l)|^2 + \sigma^2},
\end{equation}
respectively.
During operation, the fluid antenna is repositioned to maximize the chosen metric $S(l)\in\{\mathrm{SNR}(l),\mathrm{SIR}(l),\mathrm{SINR}(l)\}$. Let $l^*$ denote an (optimizing) position such that
\begin{equation}
S(l^*) = \sup_{0\leq l\leq L}\{S(l)\},
\end{equation}
and define the spatial maximum $S(l^*)\triangleq S^*$.

\subsection{An Approximation for the cdf of $S^*$}
Deriving the exact distribution of $S^*$ is mathematically intractable due to the spatial correlation between channels along the track. We therefore adopt a tractable approach that leverages the LCR of the process $S(l)$ to approximate the cdf of its spatial maximum. Specifically, the fade sojourn duration of a smooth fading process is well modeled as an exponential distribution with mean equal to the average fade duration \cite{sojourn}, and we extend this model to the spatial domain. Using this approach, we then derive an approximation for the cdf of $S^*$ as:
\begin{align}\label{s*}
F_{S^\star}(s_{\text{th}})
\;\approx\;
F_{S(l)}(s_{\text{th}})\;
\exp\!\left(
-\frac{L\,\mathrm{LCR}_{S(l)}(s_{\text{th}})}{\mathrm{F}_{S(l)}(s_{\text{th}})}
\right),
\end{align}
where $F_{S(l)}(s_{\text{th}})$ is the marginal cdf of $S(l)$ and $\mathrm{LCR}_{S(l)}(s_{\text{th}})$ is the (spatial) LCR across threshold $s_{\text{th}}$.
\begin{proof}
The expression follows by approximating the spatial fade sojourn distance as an exponential random variable with mean equal to the average fade distance. A detailed derivation is provided in Appendix~\ref{proof_CDF_s*}.
\end{proof}

Another method for characterizing the cdf of $S^*$ is developed in \cite{Psomas_ICL_2023}, given by
\begin{align}\label{bound_s*}
F_{S^*}(s_{\text{th}}) \geq F_{S(l)}(s_{\text{th}}) - L\times \mathrm{LCR}_{S(l)}(s_{\text{th}}),
\end{align}
which is asymptotically exact in the upper tail. It is interesting to note that the two methods are closely related. In particular, a first-order (Taylor) approximation of the exponential term in \eqref{s*} yields \eqref{bound_s*}. However, the approximation in \eqref{s*} is accurate for the whole cdf, whereas the lower bound in \eqref{bound_s*} is particularly accurate in the upper tail.

\subsection{Performance Analysis under Rayleigh Fading}\label{RayleighAnalysis}
We now consider the Rayleigh case, where the desired channel is $h_0(l)\sim\mathcal{CN}(0,\beta_0)$.

\subsubsection{SNR Analysis}
For Rayleigh fading, $\mathrm{SNR}(l)$ is exponentially distributed with mean
$\gamma_0 \triangleq \frac{E_{x_0}\beta_0}{\sigma^2}$.
The corresponding LCR across $s_{\text{th}}$ is available in closed form \cite{jakes}.
Substituting the exponential cdf and the Rayleigh LCR into \eqref{s*} gives
\begin{align}
F_{S^\star}^{\text{SNR}}(s_{\text{th}})
&\approx
\left(1 - e^{-s_{\text{th}}/\gamma_0}\right)
\exp\!\Bigg(
\frac{-L \, \sqrt{\frac{2b s_{\text{th}}}{\pi \gamma_0}}
\, e^{-s_{\text{th}}/\gamma_0}}
{1 - e^{-s_{\text{th}}/\gamma_0}}
\Bigg).
\label{Ray_SNR_cdf}
\end{align}

\subsubsection{SIR Analysis}
Next, consider an interference-limited regime (noise negligible), so the performance metric is $\mathrm{SIR}(l)$ in \eqref{sir}. Let
\begin{equation}
\Lambda_{0,n} \triangleq \frac{E_{x_0}\beta_0}{E_{x_n}\beta_n},\qquad n=1,\ldots,N.
\end{equation}
Using \eqref{s*}, the cdf of the spatial maximum SIR with unequal-power interferers (denoted as $\mathrm{SIR}_U$) is approximated as
\begin{align}\label{SIRU}
F_{S^\star}^{\mathrm{SIR}_U}(s_{\text{th}})
\;\approx\;
F_{S(l)}^{\mathrm{SIR}_U}(s_{\text{th}})\;
\exp\!\left(
-\frac{L\,\mathrm{LCR}_{S(l)}^{\mathrm{SIR}_U}(s_{\text{th}})}{F_{S(l)}^{\mathrm{SIR}_U}(s_{\text{th}})}
\right),
\end{align}
where
\begin{align}
F_{S(l)}^{\mathrm{SIR}_U}(s_{\text{th}}) = 1-\prod_{n=1}^{N} \frac{\Lambda_{0,n}}{\Lambda_{0,n} + s_{\text{th}}},
\end{align}
and
\begin{align}\label{LCR_SIR_rayleigh}
\mathrm{LCR}_{S(l)}^{\mathrm{SIR}_U}(s_{\text{th}})
&= \sqrt{\frac{b s_{\text{th}}}{2}}
\left( \prod_{n=1}^{N} \frac{\Lambda_{0,n}}{\Lambda_{0,n} + s_{\text{th}}} \right)
\nonumber\\&\times\sum_{n=1}^{N}\Bigg(\frac{\prod_{\substack{i=1,\, i \neq n}}^{N} \frac{\Lambda_{0,i}}{\Lambda_{0,i}-\Lambda_{0,n}}}{\sqrt{\Lambda_{0,n}}}\Bigg).
\end{align}
These expressions follow by substituting the known cdf and LCR of SIR \cite{Annavajjala_IMC_2010} into \eqref{s*}.
Note that when $E_{x_i}\beta_i=E_{x_j}\beta_j$ for any $i\neq j$, the denominator $\Lambda_{0,i}-\Lambda_{0,n}$ becomes zero. In this paper, we look at one important special case where $E_{x_i}\beta_i=E_{x}\beta$ for all $i$, and the corresponding results are given in Appendix~\ref{app:equal_power_SIR}.

\subsubsection{SINR Analysis}
We now include both noise and interference. Let $\mathrm{SINR}_1$ denote the SINR with a single interferer ($N=1$). Substituting the SINR cdf and LCR from \cite{Annavajjala_IMC_2010} into \eqref{s*} yields
\begin{align}\label{SINR1}
F_{S^\star}^{\mathrm{SINR}_1}(s_{\text{th}})
\;\approx\;
F_{S(l)}^{\mathrm{SINR}_1}(s_{\text{th}})\;
\exp\!\left(
-\frac{L\,\mathrm{LCR}_{S(l)}^{\mathrm{SINR}_1}(s_{\text{th}})}{F_{S(l)}^{\mathrm{SINR}_1}(s_{\text{th}})}
\right),
\end{align}
with
\begin{align}
F_{S(l)}^{\mathrm{SINR}_1}(s_{\text{th}})
= 1-\exp\Big(-\frac{s_{\text{th}}}{\gamma_0}\Big)\Bigg(\frac{\Lambda_{0,1}}{\Lambda_{0,1} + s_{\text{th}}}\Bigg),
\end{align}
where $\Lambda_{0,1}=\frac{E_{x_0}\beta_0}{E_{x_1}\beta_1}$, and
\begin{align}\label{LCR_SINR_rayleigh}
\mathrm{LCR}_{S(l)}^{\mathrm{SINR}_1}(s_{\text{th}})
&= \sqrt{\frac{2b s_{\text{th}}\gamma_1}{\pi\gamma_0}}
\Bigg(\frac{\Lambda_{0,1}}{\Lambda_{0,1} + s_{\text{th}}}\Bigg)
\exp \Big(\frac{1}{\gamma_1}\Big)\nonumber\\
&\times\exp\Big(-\frac{s_{\text{th}}}{\gamma_0}\Big)
\Gamma\!\left(\frac{3}{2},\frac{1}{\gamma_1}\right),
\end{align}
where $\gamma_1 \triangleq \frac{E_{x_1}\beta_1}{\sigma^2}$ and $\Gamma(u,n)$ is the upper incomplete gamma function \cite[Eq.\ (8.350-2)]{Gradshteyn_book_2007}.

Next, we consider the case with multiple interferers present ($N \ge 1$). Let $\mathrm{SINR}_U$ denote the SINR with unequal-power interferers. By substituting the cdf and the LCR of $\mathrm{SINR}_U$ from \cite{Annavajjala_IMC_2010} to \eqref{s*}, we obtain
\begin{align}\label{SINRMU}
F_{S^\star}^{\mathrm{SINR}_U}(s_{\text{th}})
\;\approx\;
F_{S(l)}^{\mathrm{SINR}_U}(s_{\text{th}})\;
\exp\!\left(
-\frac{L\,\mathrm{LCR}_{S(l)}^{\mathrm{SINR}_U}(s_{\text{th}})}{F_{S(l)}^{\mathrm{SINR}_U}(s_{\text{th}})}
\right),
\end{align}
where
\begin{align}
F_{S(l)}^{\mathrm{SINR}_U}(s_{\text{th}})=1-\exp\Big(-\frac{s_{\text{th}}}{\gamma_0}\Big)\prod_{n=1}^{N} \frac{\Lambda_{0,n}}{\Lambda_{0,n} + s_{\text{th}}},
\end{align}
and
\begin{align}\label{multiple int}
\mathrm{LCR}_{S(l)}^{\mathrm{SINR}_U}(s_{\text{th}})
&= \sqrt{\frac{2bs_{\text{th}}}{\gamma_0 \pi}} \, \exp\left(-\frac{s_{\text{th}}}{\gamma_0}\right)
\left( \prod_{n=1}^{N} \frac{\Lambda_{0,n}}{\Lambda_{0,n} + s_{\text{th}}} \right)
\nonumber\\
&\times\sum_{n=1}^{N} \delta_{n} \, \frac{\exp(W_n)}{\sqrt{W_n}} \, \Gamma\!\left(\frac{3}{2},W_{n}\right),
\end{align}
where $\mathrm{SINR}_U$ denotes the SINR with $N \ge 1$ unequal-power interferers, $\gamma_n \triangleq \frac{E_{x_n}\beta_n}{\sigma^2}$, $\Lambda_{0,n}=\frac{E_{x_0}\beta_0}{E_{x_n}\beta_n}$,
$W_n=\frac{1}{\gamma_n}$, and $\delta_{n}=\prod_{\substack{i=1,\, i \neq n}}^{N} \frac{W_{i}}{W_{i}-W_{n}}$. Note that when $\gamma_i=\gamma_j$ for any $i\neq j$, the denominator in $\delta_n$ becomes zero. In this paper, we look at one important special case where all the interferers have same power, and the corresponding results are given in Appendix~\ref{app:sinr_equal_power}.
\subsection{Performance Analysis under Ricean Fading}
We now consider a Ricean desired channel with a deterministic line-of-sight (LoS) component and a Rayleigh diffuse component, while all interferers remain Rayleigh. The desired channel is
\begin{align}\label{ric_channel}
    h_0(l) = \sqrt{\beta_0}\left(\sqrt{\frac{K}{K+1}} \, e^{-j\Phi l} + \frac{1}{\sqrt{K+1}} \, u(l)\right),
\end{align}
where $K$ is the Ricean $K$-factor, $e^{-j\Phi l}$ is the LoS phase term (linear over $[0,L]$ under standard steering-vector models), and $u(l)\sim\mathcal{CN}(0,1)$ has spatial correlation given by \eqref{correlation}.

\subsubsection{SNR Analysis}
Let the superscript $\mathrm{SNR}_{\mathrm{Ri}}$ denote the SNR of a Ricean channel. Using \eqref{ric_channel} and the general approximation \eqref{s*}, the cdf of the spatial maximum $\mathrm{SNR}_{\mathrm{Ri}}$ can be written as
\begin{align}\label{SNRRic}
F_{S^\star}^{\mathrm{SNR}_{\mathrm{Ri}}}(s_{\text{th}})
\;\approx\;
F_{S(l)}^{\mathrm{SNR}_{\mathrm{Ri}}}(s_{\text{th}})\;
\exp\!\left(
-\frac{L\,\mathrm{LCR}_{S(l)}^{\mathrm{SNR}_{\mathrm{Ri}}}(s_{\text{th}})}{F_{S(l)}^{\mathrm{SNR}_{\mathrm{Ri}}}(s_{\text{th}})}
\right),
\end{align}
The Ricean cdf and LCR expressions are taken from \cite{Cheng_ICC_2009}.
The marginal cdf is
\begin{align}
F_{S(l)}^{\mathrm{SNR}_{\mathrm{Ri}}}(s_{\text{th}}) = 1-Q_1\!\left(\sqrt{2K},\sqrt{\frac{2(K+1)s_{\text{th}}}{\gamma_0}}\right),
\end{align}
with $Q_1(\cdot,\cdot)$ the first-order Marcum $Q$-function \cite{marcum}.
The corresponding LCR is given by \eqref{lcr_snr_ric}, shown at the bottom of the next page. 
\begin{figure*}[!b]
\normalsize
\hrulefill
\begin{align}\label{lcr_snr_ric}
\mathrm{LCR}_{S(l)}^{\mathrm{SNR}_{\mathrm{Ri}}}(s_{\text{th}})
&= \frac{2(K+1)}{\pi^{\frac{3}{2}}}\sqrt{\frac{2s_{\text{th}}B}{\gamma_0}}
\exp\!\Big(-K-(K+1)\tfrac{s_{\text{th}}}{\gamma_0}\Big)
\int_{0}^{\pi/2}
\cosh\!\Big(2\sqrt{K(K+1)}\sqrt{\tfrac{s_{\text{th}}}{\gamma_0}}\cos\theta\Big)\nonumber\\
&\quad\times
\Bigg(\exp\!\Big(-\Big(\tfrac{\Phi\zeta\sin\theta}{\sqrt{2B}}\Big)^2\Big)
+ \sqrt{\tfrac{\pi}{2B}}\,\Phi\zeta\sin\theta\,\mathrm{erf}\!\Big(\tfrac{\Phi\zeta\sin\theta}{\sqrt{2B}}\Big)\Bigg)
\, d\theta.
\end{align}
\end{figure*}
For compactness we use the auxiliary parameters
\begin{equation}
\zeta \triangleq \sqrt{\frac{K}{K+1}},\qquad B \triangleq \frac{b}{K+1},
\end{equation}
which arise from the LoS amplitude and the (normalized) spatial derivative variance of the diffuse component in \eqref{ric_channel}.

\subsubsection{SIR Analysis}
We next consider the SIR scenario with a single Rayleigh interferer $h_1(l)\sim\mathcal{CN}(0,\beta_1)$. The superscript $\mathrm{SIR}_{\mathrm{Ri}}$ is used to denote the SIR when the desired channel is Ricean and the interferer is Rayleigh. Using \eqref{s*}, the cdf of the spatial maximum $\mathrm{SIR}_{\mathrm{Ri}}$ is approximated as
\begin{align}\label{SIRRic}
F_{S^\star}^{\mathrm{SIR}_{\mathrm{Ri}}}(s_{\text{th}})
\;\approx\;
F_{S(l)}^{\mathrm{SIR}_{\mathrm{Ri}}}(s_{\text{th}})\;
\exp\!\left(
-\frac{L\,\mathrm{LCR}_{S(l)}^{\mathrm{SIR}_{\mathrm{Ri}}}(s_{\text{th}})}{F_{S(l)}^{\mathrm{SIR}_{\mathrm{Ri}}}(s_{\text{th}})}
\right),
\end{align}
where
\begin{equation}\label{cdf_sir_ric}
F_{S(l)}^{\mathrm{SIR}_{\mathrm{Ri}}}(s_{\text{th}}) = \frac{f}{1+f}\,\exp\!\Big(\frac{-K}{1+f}\Big),
\end{equation}
with
\begin{equation}
    f \triangleq \frac{\beta_1 (K+1) s_{\text{th}} E_{x_1}}{\beta_0 E_{x_0}}.
\end{equation}
The LCR expression is given by \eqref{lcr_final_ric} at the top of the next page.
Define
\begin{align}
\alpha &\triangleq \sqrt{\frac{\beta_0 K}{K+1}},\\
\kappa_1 &\triangleq 2\Big(\frac{\beta_0 b}{K+1} + b\beta_1\frac{s_{\text{th}}E_{x_1}}{E_{x_0}}\Big),\\
\kappa_2 &\triangleq \frac{(K+1)E_{x_1}s_{\text{th}}}{E_{x_0}\beta_0} + \frac{1}{\beta_1},\\
\kappa_3 &\triangleq 2\sqrt{\frac{E_{x_1}s_{\text{th}}K(K+1)}{E_{x_0}\beta_0}}.
\end{align}

\begin{figure*}[!t]
\normalsize
\begin{align}
\label{lcr_final_ric}
\mathrm{LCR}_{S(l)}^{\mathrm{SIR}_{\mathrm{Ri}}}(s_{\text{th}})
&= \int_{0}^{2\pi} \sqrt{\frac{s_{\text{th}}E_{x_1}}{E_{x_0}}}\, \frac{2e^{-K}(K+1)}{\pi\beta_0\beta_1}
\Bigg(\sqrt{\frac{\kappa_1}{4\pi}}\exp\!\Big(\frac{-(\alpha \Phi\sin\theta)^2}{\kappa_1} \Big)
+ \frac{\alpha \Phi\sin\theta}{2}
\Bigg(1+\mathrm{erf}\!\Big(\frac{\alpha \Phi\sin\theta}{\sqrt{\kappa_1}}\Big)\Bigg)\Bigg)\nonumber\\
&\quad\times\Bigg(\frac{\kappa_3 \cos\theta}{4\kappa_2^2}
+ \exp\!\Big( \frac{\kappa_3^2 \cos^2 \theta}{4\kappa_2}\Big)
\frac{\sqrt{\pi}}{8\kappa_2^{\frac{5}{2}}}(\kappa_3^2 \cos^2\theta+2\kappa_2)
\Bigg(1+\mathrm{erf}\!\Big(\frac{\kappa_3 \cos\theta}{2\sqrt{\kappa_2}}\Big)\Bigg)\Bigg)
\, d\theta.
\end{align}
\hrulefill
\end{figure*}

See Appendix~\ref{Appendix:SIR_rician} for the derivation.
The numerical integration in \eqref{lcr_final_ric} arises from integrating over the (random) phase in the Ricean channel. If additional links were also Ricean, further phase integrals would appear; hence we restrict attention to the case where only the desired link is Ricean.

\textit{Note:} The Ricean model above assumes far-field propagation with linear phase. In near-field scenarios, the phase is position dependent and spherical, making the LCR position dependent. A global analysis could be performed by averaging the LCR across the antenna movement range, and the resulting average LCR can then be used in the general approximation of \eqref{s*} to analyze the near-field scenario.
\subsection{Mathematical Insights}\label{math}
We first highlight the SNR improvement afforded by mobility.
For a fixed antenna (e.g., at $l=0$), the Rayleigh SNR tail is
\begin{align}\label{uppertail}
    \mathbb{P}\big(S(0) > s_{\text{th}}\big) = \exp\!\Big(\frac{- s_{\text{th}}}{\gamma_0}\Big).
\end{align}
Using the first-order approximation \eqref{bound_s*} for a single fluid antenna yields the high-threshold asymptotic
\begin{align}
    \mathbb{P}\big(S^* > s_{\text{th}}\big)
    &\sim \exp\!\Big(\frac{- s_{\text{th}}}{\gamma_0}\Big)
    \Bigg(1 + L \sqrt{\frac{2b s_{\text{th}}}{\pi  \gamma_0}}\Bigg),
\end{align}
where $\sim$ denotes asymptotic equivalence as $s_{\text{th}}\to\infty$.
Thus, in the upper tail, mobility scales the success probability by a factor that is linear in $L$ and proportional to $\sqrt{s_{\text{th}}}$.

Applying \eqref{bound_s*} to the SINR with a single Rayleigh interferer gives
\begin{align}\label{upper_SINR}
\mathbb{P}\big(S^* > s_{\text{th}}\big)
&\sim \Bigg(\frac{\Lambda_{0,1}}{\Lambda_{0,1}+s_{\text{th}}}\Bigg)
\exp\!\Big(\frac{-s_{\text{th}}}{\gamma_0}\Big)
\nonumber\\
&\times\Bigg(1+L \sqrt{\frac{2bs_{\text{th}}\gamma_1}{\pi\gamma_0}}
\exp\!\Big(\frac{1}{\gamma_1}\Big)
\Gamma\!\Big(\frac{3}{2},\frac{1}{\gamma_1}\Big)\Bigg).
\end{align}
Again, the upper-tail probability is scaled by a term linear in $L$ and proportional to $\sqrt{s_{\text{th}}}$.

Next, we can quantify \emph{interference neutralization} by determining the length $L$ required for the (high-threshold) SINR tail in \eqref{upper_SINR} to match the fixed-antenna SNR tail in \eqref{uppertail}. Here, interference neutralization implies that the probability that the SNR exceeds $s_\text{th}$ is the same for a fluid antenna with interference and a fixed antenna with no interference. Equating \eqref{uppertail} and \eqref{upper_SINR} yields
\begin{align}\label{length}
L = \sqrt{\frac{\pi s_{\text{th}}\gamma_1}{2b\gamma_0}} \, \frac{\exp\!\big(-\frac{1}{\gamma_1}\big)}{\Gamma\!\left(\frac{3}{2},\frac{1}{\gamma_1}\right)}.
\end{align}
Hence, $L$ in \eqref{length} represents the length of track required to “neutralize” the effect of a single interferer at high thresholds. Intuitively, this means that by scanning over a track of length $L$, the fluid antenna can position itself to mitigate interference and achieve a SINR comparable to the interference-free SNR. Equation \eqref{length} also shows how $L$ scales with the interferer strength ($\gamma_1$) and the desired signal power ($\gamma_0$): strong interferers or weak desired signals require a longer track to achieve interference neutralization. In particular, the relationship between $L$ and $\gamma_0$ is an inverse square root relationship and, using the series approximation of $\Gamma(\frac{3}{2},\frac{1}{\gamma_1})$ for large values of $\gamma_1$ \cite[Eq.\ 8.7.6]{Paris_IncompleteGamma} shows that $L$ is quadratic in $\gamma_1$ for large $\gamma_1$.


\section{Multi Antenna Systems}\label{multisec}

Integrating multiple receive antennas into a fluid-antenna architecture can provide additional diversity and combining gain, thereby improving link reliability and (in general settings) capacity. In this section we study two receiver layouts in which \emph{only one component is movable}: (i) a receiver equipped with one fluid antenna assisted by a second, fixed antenna, and (ii) a rigid array of fixed antennas that is moved as a whole. The idea of one movable component is motivated by practical considerations: enabling multiple movable components in a continuous fluid antenna system introduces hardware complexity and control overhead. By contrast, allowing a single degree of movement provides a simpler and more implementable architecture. While this setup is not optimal compared to the systems with multiple movable components, it enables us to explore the gains achievable with a single source of spatial movement, serving as a meaningful intermediate step between conventional fixed arrays and fully flexible fluid antenna systems. Throughout this section we focus on the post-combining SNR in a noise-limited link under Rayleigh fading with no co-channel interference. We assume maximum ratio combining (MRC), which is optimal for maximizing the output SNR under these conditions.
\begin{figure}[t]
    \centering
    \subfloat[]{
        \includegraphics[width=\linewidth]{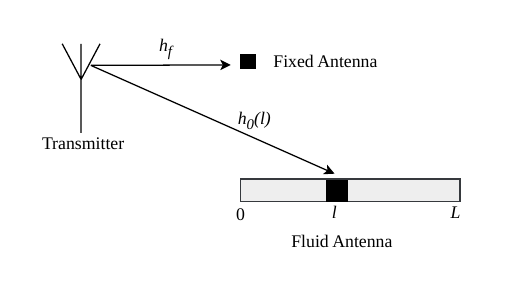}
        \label{fig:subfig_a}
    }
    
    \subfloat[]{
        \includegraphics[width=\linewidth]{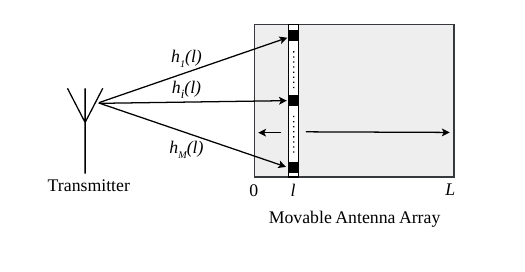}
        \label{fig:subfig_b}
    }
    \caption{Multi-antenna receiver layouts: (a) one fixed antenna and one fluid antenna, with channels denoted by $h_f$ and $h_0(l)$, respectively; (b) a rigid moving array of $M$ fixed antennas, where $h_i(l)$ denotes the channel from the transmitter to the $i$th array element.}
    \label{fig:double_column}
\end{figure}

\subsection{Single Fluid Antenna with a Fixed Antenna}
We first consider a receiver comprising one fixed antenna and one fluid antenna (Fig.~\ref{fig:subfig_a}). The fixed antenna provides a stable diversity branch, while the fluid antenna can be repositioned along a track to exploit spatial variability. The received baseband signal on the fixed branch is
\begin{align}
    y_f = h_f x_0 + n_f,
\end{align}
and the received signal on the fluid antenna branch at position $l\in[0,L]$ is
\begin{equation}
\label{received_signal_fluid}
  y(l) = h_{0}(l)x_{0} + n(l), \quad 0 \leq l \leq L,
\end{equation}
where $x_0$ is the transmitted symbol with $\mathbb{E}[|x_0|^2] = E_{x_{0}}$, $n_f$ and $n(l)$ are independent AWGN terms with zero mean and variance $\sigma^2$, $h_f\sim\mathcal{CN}(0,\beta_f)$ is the channel to the fixed antenna, and $h_{0}(l)\sim\mathcal{CN}(0,\beta_0)$ is the channel to the fluid antenna at position $l$. We assume the fixed antenna and the fluid antenna reference position are separated by at least half a wavelength so that the two branches are effectively uncorrelated. Along the fluid antenna track, we assume the spatial correlation model satisfies \eqref{correlation2}.

With MRC, the instantaneous post-combining SNR when the fluid antenna is at $l$ is
\begin{equation}\label{snr_fixed}
S(l) = \frac{E_{x_{0}}}{\sigma^2}\big(|h_f|^2 + |h_0(l)|^2\big).
\end{equation}
The FA is positioned to maximize $S(l)$ over $l\in[0,L]$. Denoting the spatial maximizer by $l^*$ and the corresponding maximum by $S^*\triangleq S(l^*)$, we approximate the cdf of $S^*$ using the general LCR-based approximation in \eqref{s*}.

\paragraph{Unequal average branch powers ($\beta_0\neq\beta_f$)}
Here we consider the general case of unequal average powers. We use the superscript $\mathrm{FF,uneq}$ to denote the single fluid antenna with a fixed antenna with unequal average branch powers. Using this notation, an approximation to the cdf of the spatial maximum SNR can be written as
\begin{align}\label{onefixed}
F_{S^\star}^{\mathrm{FF},\mathrm{uneq}}(s_{\text{th}})
\;\approx\;
F_{S(l)}^{\mathrm{FF},\mathrm{uneq}}(s_{\text{th}})\;
\exp\!\left(
-\frac{L\,\mathrm{LCR}_{S(l)}^{\mathrm{FF},\mathrm{uneq}}(s_{\text{th}})}{F_{S(l)}^{\mathrm{FF},\mathrm{uneq}}(s_{\text{th}})}
\right),
\end{align}
where the marginal cdf is
\begin{align}\label{cdf_fixed_fluid}
   F_{S(l)}^{\mathrm{FF},\mathrm{uneq}}(s_{\text{th}})= 1- \frac{1}{\varsigma}
   \Bigg(
   \frac{\exp\!\big(-\frac{\sigma^2 s_{\text{th}}}{E_{x_{0}}\beta_f}\big)}{\beta_{0}}
   -\frac{\exp\!\big(-\frac{\sigma^2 s_{\text{th}}}{E_{x_{0}}\beta_0}\big)}{\beta_f}
   \Bigg),
\end{align}
and the corresponding LCR is
\begin{align}\label{lcr1fix}
\mathrm{LCR}_{S(l)}^{\mathrm{FF},\mathrm{uneq}}(s_{\text{th}})
&= \sqrt{\frac{2 b \beta_0}{\pi}}\,\frac{\exp\!\big(-\frac{\sigma^2 s_{\text{th}}}{E_{x_{0}}\beta_f}\big)}{\beta_0 \beta_f}
\frac{\gamma\!\left(\frac{3}{2},\,\varsigma\frac{\sigma^2 s_{\text{th}}}{E_{x_{0}}}\right)}{\varsigma^{\frac{3}{2}}},
\end{align}
with $\varsigma = \frac{1}{\beta_0}-\frac{1}{\beta_f}$ and where $\gamma(u,n)$ is the lower incomplete gamma function \cite[Eq.~(8.350-1)]{Gradshteyn_book_2007}.

\paragraph{Equal average branch powers ($\beta_0=\beta_f=\beta$)}
The equal-power case is practically important when the fixed and fluid antennas are closely spaced, but it cannot be obtained by directly substituting $\beta_0=\beta_f$ into \eqref{onefixed}--\eqref{lcr1fix} due to zero divisors. We denote this scenario with the superscript $\mathrm{FF,eq}$. For $\beta_f=\beta_0=\beta$, the marginal post-combining SNR in \eqref{snr_fixed} is gamma distributed with shape~2, and applying \eqref{s*} yields
\begin{align}\label{FF_EqP}
F_{S^\star}^{\mathrm{FF,eq}}(s_{\text{th}})
\;\approx\;
\gamma\!\left(2,\frac{\sigma^2 s_{\text{th}}}{E_{x_{0}} \beta}\right)
\exp\!\left(
-\frac{L\,\mathrm{LCR}_{S(l)}^{\mathrm{FF,eq}}(s_{\text{th}})}{\gamma\!\left(2,\frac{\sigma^2 s_{\text{th}}}{E_{x_{0}} \beta}\right)}
\right),
\end{align}
where the required LCR is
\begin{align}\label{lcrfix1same}
\mathrm{LCR}_{S(l)}^{\mathrm{FF,eq}}(s_{\text{th}})
= \frac{2}{3}\,\sqrt{\frac{2 b}{\pi}}
\exp\!\Big(-\frac{\sigma^2 s_{\text{th}}}{E_{x_{0}}\beta}\Big)
\left(\frac{\sigma^2 s_{\text{th}}}{E_{x_{0}}\beta}\right)^{\frac{3}{2}}.
\end{align}
See Appendix~\ref{appendix:cdf_onefixed_onefluid} for detailed derivations.

\subsection{Moving array of fixed antennas}
We next consider a receiver in which a rigid array of $M$ fixed antennas is embedded within a fluid-antenna structure (Fig.~\ref{fig:subfig_b}). The array elements maintain fixed relative positions, while the \emph{entire array} is translated along a track of length $L$ and is positioned to maximize the post-combining SNR. The separation between adjacent antennas is denoted by $\Delta$ (measured in wavelengths). With MRC, the instantaneous SNR at array position $l\in[0,L]$ can be written as
\begin{equation}\label{snr_array}
S(l) = \frac{E_{x_{0}}}{\sigma^2}\, \mathbf{h}^{\mathrm{H}}(l)\mathbf{h}(l)
= \frac{E_{x_{0}}}{\sigma^2}\sum_{i=1}^{M} |h_i(l)|^2,
\end{equation}
where $\mathbf{h}(l) = \left[ h_1(l), \, h_2(l), \, \cdots, \, h_M(l) \right]^T$ is the channel vector at location $l$. Since the array is compact, we assume equal average channel powers $\mathbb{E}[|h_i(l)|^2]=\beta$ and $h_i(l)\sim\mathcal{CN}(0,\beta)$ for all $i\in\{1,\ldots,M\}$.

\paragraph{Widely spaced array elements (approximately uncorrelated branches)}
When the inter-element spacing $\Delta$ is sufficiently large, the antenna branches can be well-modelled as uncorrelated. In this case the moving array behaves as an (approximately) independent-branch MRC receiver whose \emph{position-dependent} fading remains spatially correlated along the track. The basic approximation in \eqref{s*} therefore applies, requiring only the marginal cdf and spatial LCR of $S(l)$. For the two-branch case (used for comparison in the numerical section), the cdf and LCR of $S(l)$ are available in \cite{Yacoub_ITVT_2001}. We use the superscript $\mathrm{arr_{ind}}$ to denote the array with independent elements. Substituting these into \eqref{s*} yields
\begin{align}\label{array_ind}
F_{S^\star}^{\mathrm{arr_{ind}}}(s_{\text{th}})
\;\approx\;
\gamma\!\left(2,\frac{\sigma^2 s_{\text{th}}}{E_{x_{0}} \beta}\right)
\exp\!\left(
-\frac{L\,\mathrm{LCR}_{S(l)}^{\mathrm{arr_{ind}}}(s_{\text{th}})}{\gamma\!\left(2,\frac{\sigma^2 s_{\text{th}}}{E_{x_{0}} \beta}\right)}
\right),
\end{align}
where
\begin{align}\label{lcr_uncorrelated_array}
\mathrm{LCR}_{S(l)}^{\mathrm{arr_{ind}}}(s_{\text{th}})
= \sqrt{\frac{2b}{\pi}}
\left(\frac{\sigma^2 s_{\text{th}}}{E_{x_{0}}\beta}\right)^{\frac{3}{2}}
\exp\!\Big(-\frac{\sigma^2 s_{\text{th}}}{E_{x_{0}}\beta} \Big).
\end{align}

\paragraph{Compact array elements (correlated branches)}
When $\Delta<\tfrac{1}{2}$, the array elements are significantly correlated and the required LCR is more challenging to obtain. Most classical LCR derivations rely on separability between temporal and spatial variations; here, both the inter-element dependence and the dependence across array locations arise from \emph{spatial} correlation, so separability does not hold and the two dimensions are intrinsically coupled. We therefore develop a dedicated analysis (Appendix~\ref{Appendxi:LCR_array_de}) for a correlated two-element moving array. We use the superscript $\mathrm{arr_{corr}}$ to denote the correlated two-element array. Applying \eqref{s*} gives
\begin{align}\label{corr_array}
F_{S^\star}^{\mathrm{arr_{corr}}}(s_{\text{th}})
\;\approx\;
F_{S(l)}^{\mathrm{arr_{corr}}}(s_{\text{th}})\;
\exp\!\left(
-\frac{L\,\mathrm{LCR}_{S(l)}^{\mathrm{arr_{corr}}}(s_{\text{th}})}{F_{S(l)}^{\mathrm{arr_{corr}}}(s_{\text{th}})}
\right),
\end{align}
where the marginal cdf is
\begin{align}
    F_{S(l)}^{\mathrm{arr_{corr}}}(s_{\text{th}}) &= 1+\left(\frac{1-J}{2J}\right)\exp\left(-\frac{(1+J)\sigma^2 s_{\text{th}}}{ E_{x_0}\beta}\right)\nonumber\\
    &\quad-\left(\frac{1+J}{2J}\right)\exp\left(-\frac{(1-J)\sigma^2 s_{\text{th}}}{ E_{x_0}\beta}\right),
\end{align}
and the corresponding LCR is
\begin{align}\label{corr_array_lcr}
\mathrm{LCR}_{S(l)}^{\mathrm{arr_{corr}}}(s_{\text{th}})
= \sqrt{\frac{c_1(1-J^2)}{2\pi J^3}}\,\exp\left(\frac{-(c_1+Jb)s_{\text{th}}}{c_1(1-J^2)c_0}\right) \nonumber\\
\quad\times \Big(e^{x_2} \big(F(\sqrt{x_2}) - \sqrt{x_2}\big) - e^{x_1} \big(F(\sqrt{x_1}) - \sqrt{x_1}\big)\Big),
\end{align}
where $F(\cdot)$ is the Dawson integral \cite[Chap.~6]{AS}, $c_0 = \frac{E_{x_{0}} \beta}{\sigma^2}$,
$J=J_0(2\pi\Delta)$, $c_1=\frac{\pi}{\Delta}J_1(2\pi\Delta)$,
$x_1 = \frac{J(b+c_1)s_{\text{th}}}{(1-J^2)c_1c_0}$,
$x_2 = \frac{J(b-c_1)s_{\text{th}}}{(1-J^2)c_1c_0}$,
and $J_1(\cdot)$ is the first-order Bessel function of the first kind \cite[Eq.~(8.402)]{Gradshteyn_book_2007}.
The expression in \eqref{corr_array_lcr} remains well-defined for all $\Delta$, even though $J = J_0(2\pi \Delta)$ can vanish for certain spacings. When $J \to 0$, both $x_1$ and $x_2$ approach zero. Using the series expansion of the Dawson integral for small arguments, we have
\begin{equation}
F(\sqrt{x_i}) - \sqrt{x_i} = -\frac{2}{3} x_i^{3/2} + o(x_i^{3/2}), \quad x_i \to 0, \; i=1,2.
\end{equation}
Hence, the bracketed expression on the second line of \eqref{corr_array_lcr} has a leading term proportional to $J^{3/2}$, which cancels the $J^{3/2}$ factor in the denominator, ensuring that the LCR remains finite as $J \to 0$. Finally, \eqref{corr_array} is obtained by combining \eqref{s*} with the above LCR and the cdf of a hypoexponential distribution \cite{Norman_book_1971}. Note that, while we present the analysis for a two-element array here, the proposed approximation framework can be extended to larger arrays.

\section{Numerical Results}\label{numres}
In this section, we validate the analytical approximations developed throughout the paper using Monte Carlo simulations, and we quantify two key benefits of FASs: SNR enhancement through spatial selection and interference neutralization through spatial adaptation. Unless otherwise stated, we adopt the Jakes' correlation model so that the local correlation expansion parameter is $b=\pi^2$, and we approximate continuous positioning in simulation by sampling the track with spatial resolution $\tau=10^{-3}$ wavelengths. All distances and movement lengths are measured in wavelengths.

\subsection{Single Fluid Antenna System}
\begin{figure}[tb]\centering
  \includegraphics[width=\linewidth]{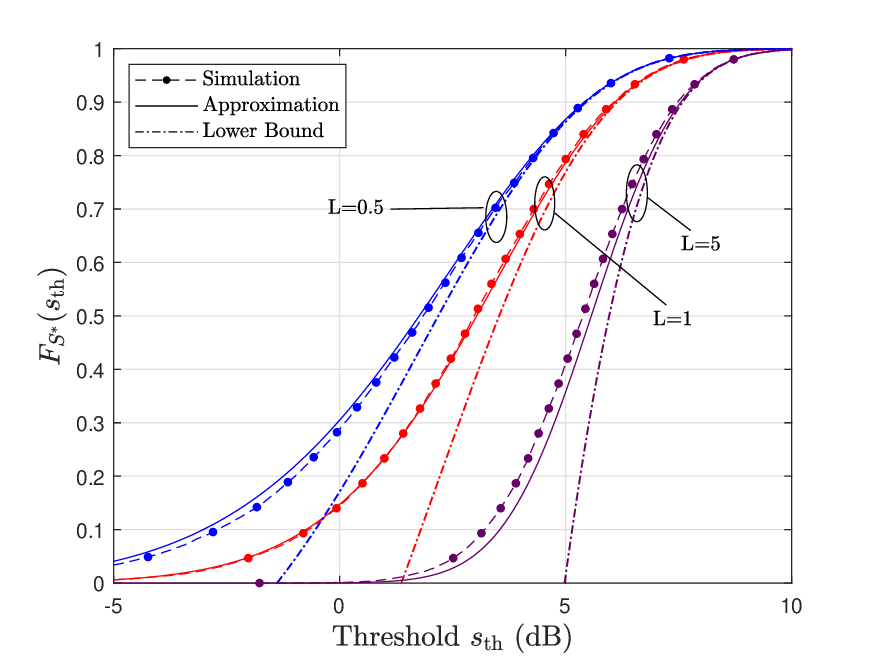}
    \caption{SNR cdf for a single fluid antenna under Rayleigh fading: analytical approximation \eqref{Ray_SNR_cdf}, first-order lower bound \eqref{bound_s*}, and Monte Carlo simulation results for $L\in\{0.5,1,5\}$.}
    \captionsetup{justification=centering}
    \label{snr_pic}
\end{figure}

We first consider the single fluid antenna receiver. Fig.~\ref{snr_pic} shows the cdf of the received SNR for a single fluid antenna under Rayleigh fading. For clarity, we normalize the average SNR to $\gamma_0=1$. Analytical approximations are obtained from \eqref{Ray_SNR_cdf} and the corresponding first-order lower bounds are obtained from \eqref{bound_s*}, where
$F_{S(l)}(s_{\text{th}})=1-e^{-s_{\text{th}}/\gamma_0}$ and
$\mathrm{LCR}_{S(l)}(s_{\text{th}})=\sqrt{\tfrac{2 b s_{\text{th}}}{\pi\gamma_0}}\,\exp\!\left(-\tfrac{s_{\text{th}}}{\gamma_0}\right)$.
Results are shown for three movement lengths, $L\in\{0.5,1,5\}$, chosen to illustrate how the performance varies with the available length, covering small to moderate ranges that are practically feasible. As expected, performance improves with $L$ because a longer track exposes the receiver to greater spatial variability and therefore increases the probability of finding a favorable channel realization. For all three lengths, the approximation is highly accurate in the high-threshold (low-outage) regime, which is the regime of primary interest for reliability analysis. Notably, for $L=1$ the approximation is accurate across essentially the entire threshold range. For $L=0.5$ the approximation slightly overestimates the simulated cdf at low thresholds, whereas for $L=5$ it slightly underestimates the cdf in that region. The lower bound remains below the simulated curves across the full range, as expected, and becomes tight in the high-threshold regime.

\begin{figure}[tb]\centering
  \includegraphics[width=\linewidth]{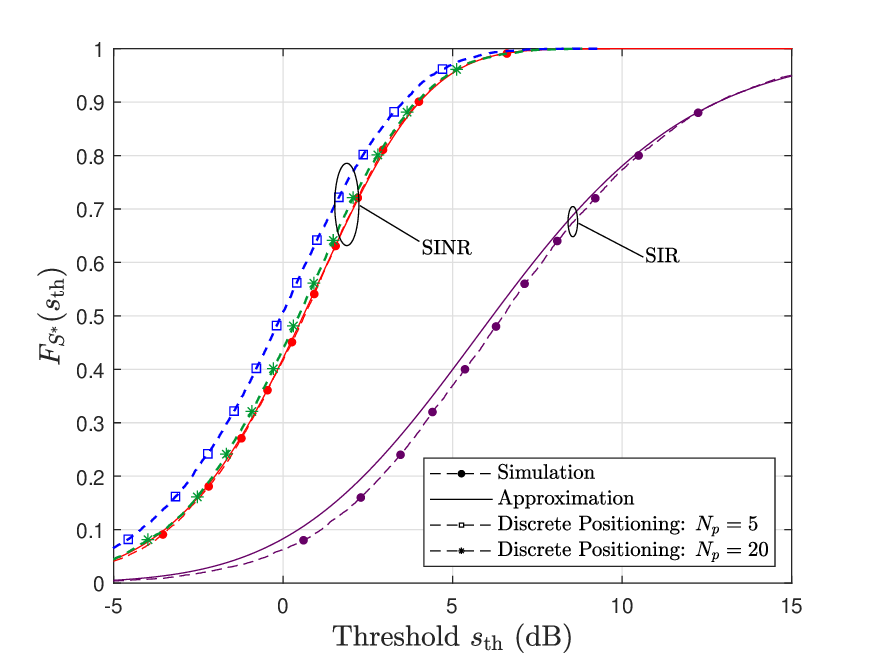}
    \caption{SINR and SIR cdfs for a single fluid antenna of length $L=1$ under Rayleigh fading with two interferers ($\Lambda_{0,1}=0.6$, $\Lambda_{0,2}=0.4$). Analytical results correspond to continuous positioning; simulations are shown for continuous and discrete positioning with $N_p\in\{5,20\}$.}
    \captionsetup{justification=centering}
    \label{sinr_pic}
\end{figure}

Fig.~\ref{sinr_pic} presents the cdf of the received SINR and SIR for a single fluid antenna under Rayleigh fading with $N=2$ interferers. The fluid antenna length is $L=1$ and the relative (desired-to-interferer) power ratios are set to $\Lambda_{0,1}=0.6$ and $\Lambda_{0,2}=0.4$. The analytical approximations are obtained from \eqref{SIRU} (SIR) and \eqref{SINRMU} (SINR). The SINR approximation tracks the simulation closely over the full threshold range, whereas the SIR approximation is slightly optimistic at low thresholds. To assess the practical impact of discrete port constraints, we also simulate discrete positioning with $N_p\in\{5,20\}$ ports. With $N_p=20$ the discrete-port performance is essentially indistinguishable from the continuous-positioning benchmark, confirming that sufficiently fine port discretization can closely approach the continuous upper bound. More generally, the continuous-positioning results provide an upper bound on any discrete-port implementation of the same physical length.

\begin{figure}[tb]\centering
  \includegraphics[width=\linewidth]{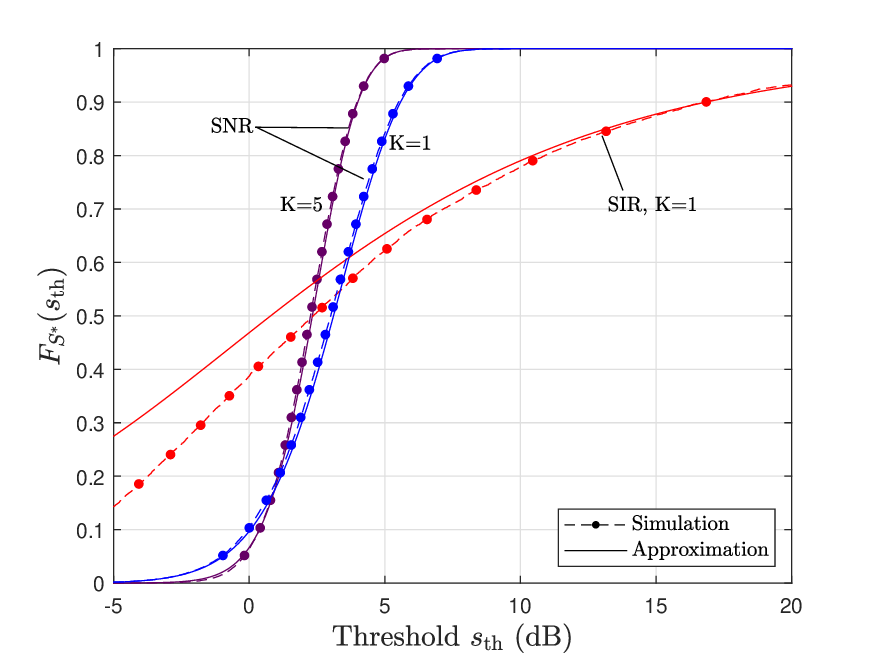}\vspace{-0mm}
    \caption{SNR and SIR cdfs for a single fluid antenna of length $L=1$ under Ricean fading on the desired channel with Rayleigh interference. SNR is shown for $K=1$ and $K=5$, while SIR is shown for $K=1$.}
    \captionsetup{justification=centering}
    \label{snr_ric}
\end{figure}

We next consider Ricean fading on the desired link, with Rayleigh interference. For the Ricean channel, we set $\Phi=2\pi$ where $\Phi$ is the LoS phase term $e^{-j\Phi l}$ introduced in the Ricean channel model. Fig.~\ref{snr_ric} shows cdfs for the received SNR and SIR of a single fluid antenna with $L=1$. For SNR, the approximation is generated from \eqref{SNRRic} with $\gamma_0=1$ and two Ricean factors $K\in\{1,5\}$. For SIR, the approximation is generated from \eqref{SIRRic} with $\beta_1=10$, $E_{x_1}=1$, $\beta_0=1$, and $E_{x_0}=1$ (and $K=1$). The SNR approximation is accurate across the full threshold range, while the SIR approximation is most accurate in the high-threshold region and slightly overestimates the cdf at low thresholds. The mismatch in SIR occurs because the SIR process is inherently less smooth. For SNR and SINR, the denominator includes a noise term, which makes the process smooth. In contrast, SIR has only interference terms in the denominator, and hence SIR fluctuates rapidly becoming heavy tailed. Because of this, the usual assumption that fade sojourn duration is exponentially distributed is less accurate for SIR, causing the approximation to overestimate the cdf at low thresholds. Two qualitative behaviors are evident. First, compared to SNR, the SIR can exhibit larger peak values because the fluid antenna may locate positions where interference is locally weak; however, the lower tail can be worse because the interference may remain non-negligible over the entire track. Second, increasing $K$ shifts the SNR cdf leftwards: as the LoS component strengthens, fading variability decreases, reducing the selection gain available from movement.

\begin{figure}[tb]\centering
  \includegraphics[width=\linewidth]{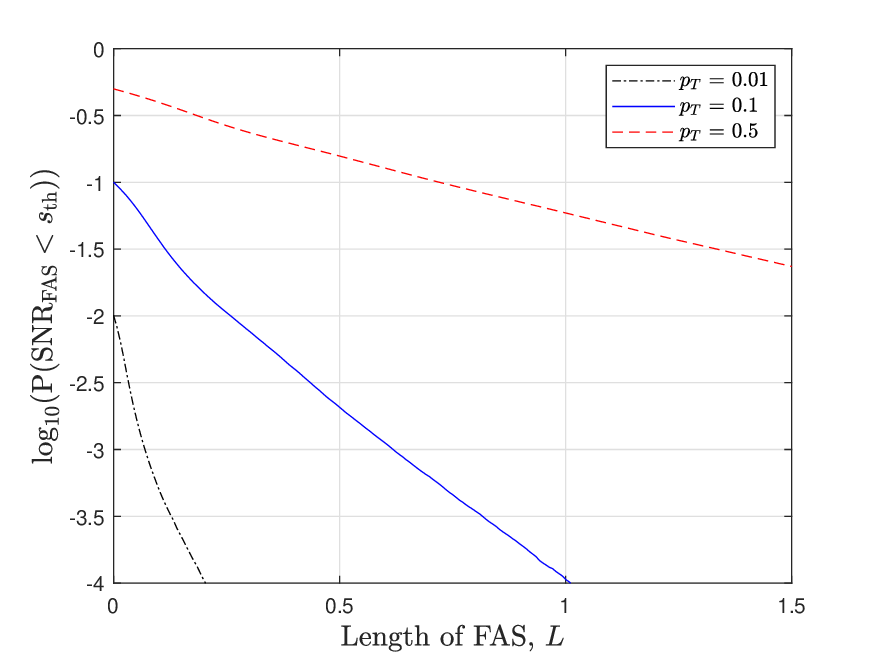}
    \caption{Impact of movable length on the lower tail of the SNR distribution for a single fluid antenna system without interference, shown for $p_T\in\{0.01,0.1,0.5\}$.}
    \captionsetup{justification=centering}
    \label{tailreduction}
\end{figure}

Fig.~\ref{tailreduction} highlights the impact of movement length on the \emph{lower tail} of the SNR distribution in an interference-free Rayleigh channel. For each target probability $p_T$, we define the threshold $s_{\text{th}}$ via the fixed-antenna relation $\mathbb{P}(\mathrm{SNR}_{\text{fixed}}<s_{\text{th}})=p_T$, where $\mathrm{SNR}_\text{fixed}$ is the SNR of a fixed antenna. We then plot $\mathbb{P}(\mathrm{SNR}_{\text{FAS}}<s_{\text{th}})$ as a function of $L$ where $\mathrm{SNR}_\text{FAS}$ is the SNR of a single fluid antenna. The results show that modest movement lengths can produce dramatic outage reductions when $p_T$ is small. For example, at $p_T=0.1$ the outage probability drops by roughly one order of magnitude at $L\approx 0.25$, two orders at $L\approx 0.62$, and three orders at $L\approx 1$. The benefit becomes even more pronounced at more stringent targets (e.g., $p_T=0.01$), whereas for median-level performance ($p_T=0.5$) the gains are comparatively modest.

\begin{figure}[tb]\centering
  \includegraphics[width=\linewidth]{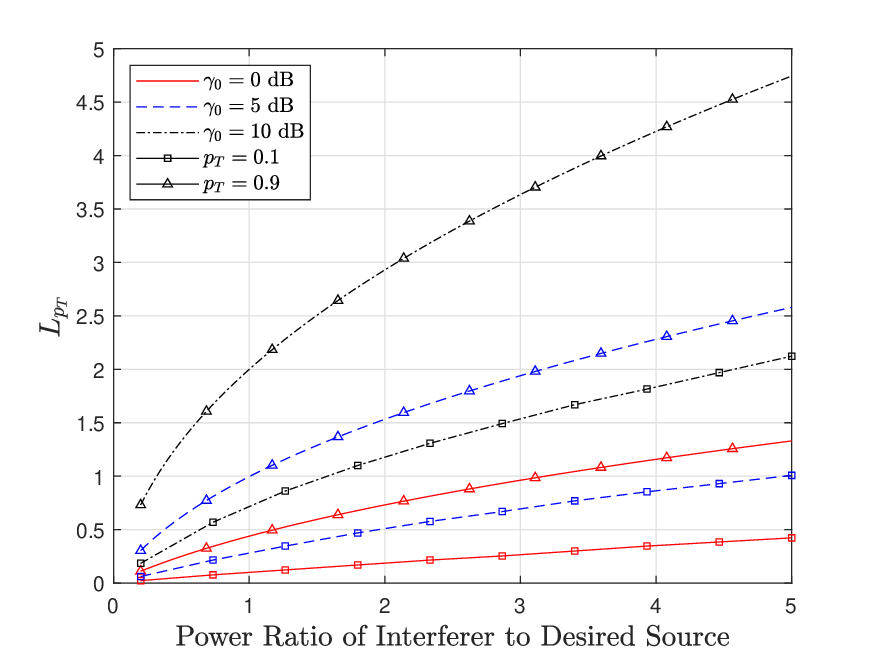}
    \caption{Movable length required to neutralize a single interferer for target probabilities $p_T\in\{0.1,0.9\}$, shown versus interferer-to-desired power ratio for $\gamma_0\in\{0,5,10\}$ dB.}
    \captionsetup{justification=centering}
    \label{intremoval}
\end{figure}

As discussed in Sec.~\ref{math}, movement can also be used to \emph{neutralize} the impact of interference at a specified performance level. For a target probability $p_T$, we define the FAS length required to neutralize the interference effect as $L_{p_T}$. Specifically, $L_{p_T}$ is chosen such that 
$\mathbb{P}(\mathrm{SNR}_{\text{fixed}}<s_{\text{th}})=p_T=\mathbb{P}(\mathrm{SINR}_{\text{FAS}}<s_{\text{th}})$,
where $\mathrm{SINR}_{\text{FAS}}$ corresponds to the SINR of a FAS. Fig.~\ref{intremoval} plots the required length as a function of the interferer-to-desired power ratio for three desired-link SNR values, $\gamma_0\in\{0,5,10\}$ dB. Results are shown for the upper tail ($p_T=0.9$), where the asymptotic expression \eqref{length} applies, and for the lower tail ($p_T=0.1$), where we report simulation-based values. The required length increases with both interference strength and the desired-link SNR, and the growth is nonlinear, consistent with \eqref{length}. Importantly, matching outage performance at the lower tail typically requires substantially shorter lengths than matching upper-tail performance. This is because lower tail performance is easier to achieve for two reasons: first, the target probability is small, so only a small number of positions need to fall below the threshold; second, the corresponding SINR threshold is lower, meaning that most antenna positions naturally exceed it. In contrast, upper tail thresholds correspond to high-SINR positions, which require scanning a longer track to achieve the desired probability. For example, at $\gamma_0=5$ dB and $p_T=0.1$, a movement length of one wavelength can neutralize an interferer that is approximately five times stronger than the desired source.

\subsection{Multi Antenna Systems}
\begin{figure}[tb]\centering
  \includegraphics[width=\linewidth]{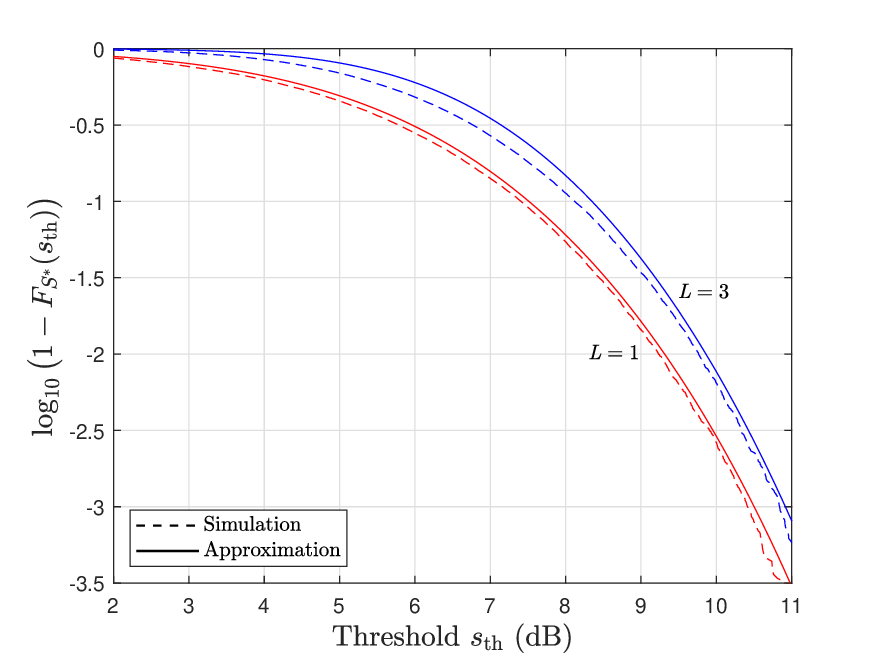}
    \caption{Log-scale ccdf of the SNR for a receiver comprising one fixed antenna and one fluid antenna, shown for $L=1$ and $L=3$.}
    \captionsetup{justification=centering}
    \label{oneifxed}
\end{figure}

We now consider the multi-antenna layouts of Sec.~\ref{multisec}. Based on the equal-power analysis in \eqref{FF_EqP}, Fig.~\ref{oneifxed} shows the complementary cdf (ccdf) of the post-combining SNR for the receiver comprising one fixed antenna and one fluid antenna, plotted on a logarithmic scale to emphasize the tail behavior. We set $\beta=1$ and $E_{x_0}=1$, and we consider two fluid antenna lengths, $L\in\{1,3\}$. The approximation is accurate across the full SNR range for $L=1$, while for $L=3$ it becomes slightly optimistic in the mid-SNR region. As in the single-antenna case, increasing $L$ improves performance by increasing spatial variability in the movable branch.

\begin{figure}[tb]\centering
  \includegraphics[width=\linewidth]{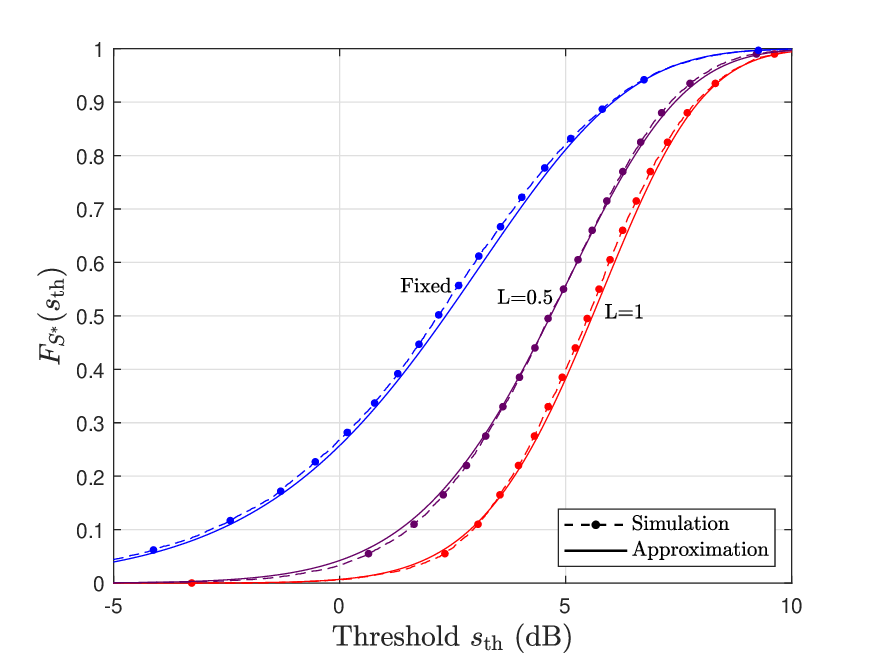}
    \caption{SNR cdf of a correlated two-element moving array, shown for a fixed array and for movement lengths $L=0.5$ and $L=1$.}
    \captionsetup{justification=centering}
    \label{array}
\end{figure}

Next, Fig.~\ref{array} shows the SNR cdf for a correlated two-element \emph{moving} array. We consider a compact two-element array and evaluate two movement ranges, $L\in\{0.5,1\}$. For comparison, we also include the cdf of the corresponding fixed array. The analytical approximation is obtained from \eqref{corr_array} and is accurate across the full threshold range for the considered lengths. Notably, even a half-wavelength movement provides a substantial gain, dramatically reducing the outage probabilities relative to a fixed array.

\begin{figure}[tb]\centering
  \includegraphics[width=\linewidth]{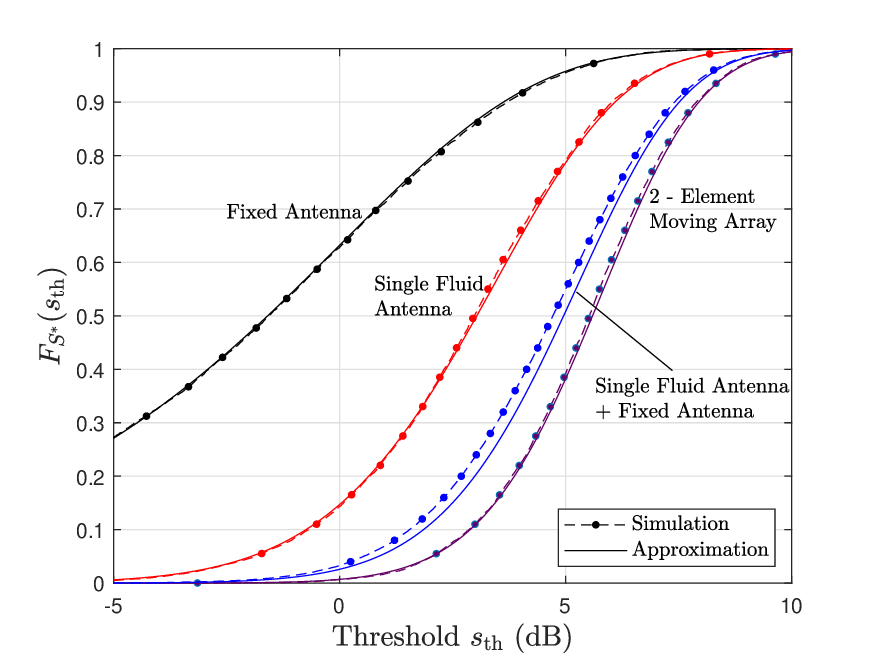}
    \caption{SNR cdf comparison (Rayleigh fading, no interference): (a) fixed antenna, (b) single fluid antenna, (c) single fluid antenna with a fixed antenna, and (d) correlated two-element moving array.}
    \captionsetup{justification=centering}
    \label{comparison}
\end{figure}

Finally, Fig.~\ref{comparison} compares the SNR cdfs (Rayleigh fading, no interference) for all three receiver layouts considered in this paper against a fixed single-antenna baseline. We set $\beta_0=1$, $\sigma^2=1$, $E_{x_0}=1$, and movement length $L=1$ for the movable components. As expected, performance improves as the receiver has access to greater spatial variability: the fixed antenna exhibits the worst SNR performance, followed by the single fluid antenna, then the fixed-plus-fluid receiver, while the correlated two-element moving array provides the best performance. Although the upper-tail curves begin to converge at high thresholds, each additional source of spatial variability still yields meaningful gains in outage probability.

Overall, the simulations confirm that the LCR-based approximations accurately characterise the cdf behaviour in the regimes of practical interest, and they highlight how increased spatial variability---through longer movement ranges, additional diversity branches, or movable arrays---translates into substantial reliability improvements. Moreover, the interference-neutralization results demonstrate that relatively modest movement lengths can neutralize strong interferers at outage-relevant operating points.

\section{Conclusion}\label{conc}

This paper investigated three receiver layouts for FASs employing single- and multi-antenna architectures. Assuming \emph{continuous positioning} over a finite track and \emph{spatial correlation} models, we addressed the challenging problem of characterising the distribution of optimised link-quality metrics. In particular, we developed analytically tractable, asymptotically accurate approximations for the cdfs of the spatial maxima of SNR, SIR, and SINR under Rayleigh fading, and extended the treatment to Ricean desired channels (with Rayleigh interference) to capture line-of-sight effects.

A central outcome is a unified analysis methodology that leverages the LCR of the underlying spatial process to approximate the cdf of the optimised performance metric attained after repositioning. The resulting expressions show strong agreement with Monte Carlo simulations in the operating regimes of interest, and they enable clear physical interpretation of how spatial mobility translates into performance gains.

The numerical results highlight several key insights. First, movement length is a powerful new degree of freedom: even modest increases in the available track length can yield dramatic reductions in outage probability, with improvements of several orders of magnitude achievable with movement on the order of one wavelength in representative scenarios. Second, FAS can mitigate co-channel interference by selecting locations that simultaneously enhance the desired signal and suppress interferers; the derived scaling laws and length conditions quantify the movement required to effectively neutralize strong interferers. Third, additional sources of spatial variability, such as adding a fixed diversity branch alongside a fluid antenna or translating a compact array as a whole, provide further gains beyond the single-antenna FAS and can substantially strengthen reliability.

Overall, the presented analytical framework and results demonstrate that fluid antenna systems offer a compelling pathway to exploit spatial variability within compact apertures, enabling substantial reliability improvements and effective interference suppression with relatively modest physical movement.

\appendices

\section{An approximation for the cdf of $S^\star$}\label{proof_CDF_s*}
Let $N^{+}_{s_{\text{th}}}(0,L)$ denote the number of \emph{upcrossings} of the level $s_{\text{th}}$ by the spatial process $S(l)$ over the interval $[0,L]$.
Observe that the event $\{S^{\star}\le s_{\text{th}}\}$ is equivalent to the event that the process starts below the threshold and never upcrosses it over the track, i.e.,
\begin{align}
\{S^{\star}\le s_{\text{th}}\}
\;\Leftrightarrow\;
\{S(0)\le s_{\text{th}},\; N^{+}_{s_{\text{th}}}(0,L)=0\}.
\end{align}
Hence, the cdf of $S^{\star}$ can be written exactly as
\begin{align}
F_{S^{\star}}(s_{\text{th}})
&= \mathbb{P}\!\left(S(0)\le s_{\text{th}},\; N^{+}_{s_{\text{th}}}(0,L)=0\right) \\
&= F_{S(l)}(s_{\text{th}})\;\mathbb{P}\!\left(N^{+}_{s_{\text{th}}}(0,L)=0\;\middle|\; S(0)\le s_{\text{th}}\right),
\label{eq:exact_identity}
\end{align}
where $F_{S(l)}(\cdot)$ denotes the marginal cdf of $S(l)$ (which is independent of $l$ under stationarity).

\begin{figure}[t]\centering
  \includegraphics[width=\linewidth]{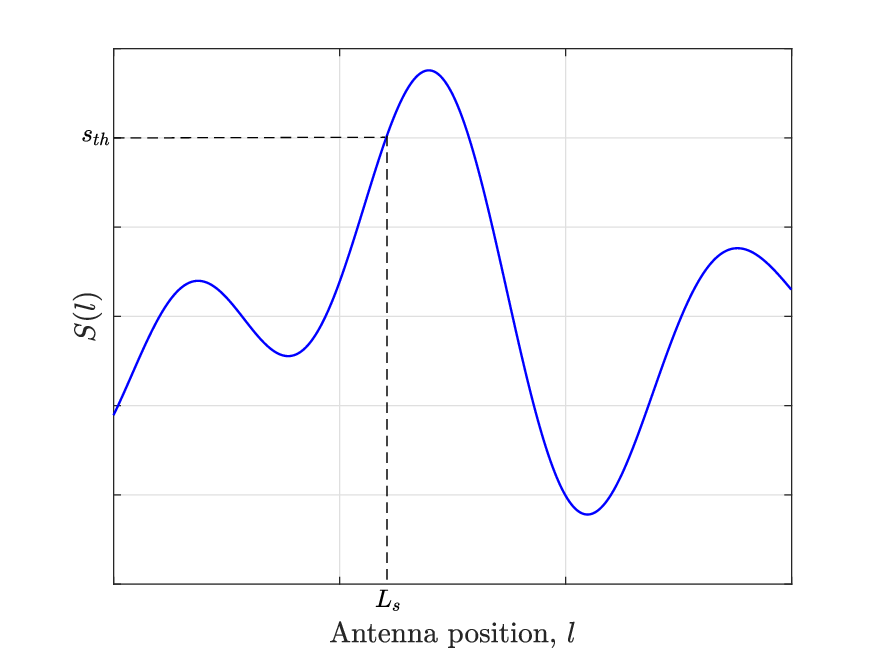}
    \caption{The process $S(l)$ and the sojourn distance $L_s$ below the threshold $s_{\text{th}}$.}
    \captionsetup{justification=centering}
    \label{sojourn}
\end{figure}

\paragraph{Sojourn-distance approximation.}
For smooth fading processes, it is common to approximate the \emph{fade sojourn time} as exponential with mean equal to the average fade duration (AFD) \cite{sojourn}. In the spatial domain, we make the analogous approximation that the \emph{fade sojourn distance}, denoted as $L_s$ (the distance the process spends below the threshold before leaving the region $S(l)\le s_{\text{th}}$; see Fig.~\ref{sojourn}) is exponential with mean equal to the \emph{average fade distance} $\mathrm{AFD}_{S(l)}(s_{\text{th}})$, i.e.,
\begin{align}
L_s \sim \mathrm{Exp}\!\left(\frac{1}{\mathrm{AFD}_{S(l)}(s_{\text{th}})}\right),
\end{align}
where $\mathbb{E}[L_s]=\mathrm{AFD}_{S(l)}(s_{\text{th}})$. Equivalently, the indicator process $\mathbb{I}_{\{S(l)\le s_{\text{th}}\}}$ is approximated by a two-state continuous-time Markov chain.
Under this approximation,
\begin{align}
\mathbb{P}\!\left(N^{+}_{s_{\text{th}}}(0,L)=0\;\middle|\;S(0)\le s_{\text{th}}\right)
&= \mathbb{P}(L_s > L)\nonumber\\
&\approx \exp\!\Big(-L/\mathrm{AFD}_{S(l)}(s_{\text{th}})\Big).
\label{eq:no_upcross_exp}
\end{align}

\paragraph{Relating AFD and LCR}
For a stationary process, the average fade distance satisfies the standard identity \cite{Zlatanov_ICL_2008}
\begin{align}
\mathrm{AFD}_{S(l)}(s_{\text{th}})
\;=\;\frac{F_{S(l)}(s_{\text{th}})}{\mathrm{LCR}_{S(l)}(s_{\text{th}})},
\end{align}
where $\mathrm{LCR}_{S(l)}(s_{\text{th}})$ is the (spatial) level crossing rate across $s_{\text{th}}$.
Substituting this into \eqref{eq:no_upcross_exp} yields
\begin{align}\label{p_app}
\mathbb{P}\!\left(N^{+}_{s_{\text{th}}}(0,L)=0\;\middle|\;S(0)\le s_{\text{th}}\right)
\;\approx\;
\exp\!\left(
-\frac{L\,\mathrm{LCR}_{S(l)}(s_{\text{th}})}{F_{S(l)}(s_{\text{th}})}
\right).
\end{align}
Finally, substituting \eqref{p_app} into \eqref{eq:exact_identity} yields \eqref{s*}.

\section{cdf of SIR for Equal-Power Interferers}\label{app:equal_power_SIR}
The superscript $\mathrm{SIR}_{\mathrm{E}}$ is used to denote the SIR for Rayleigh fading with $N$ \emph{equal-power} interferers. The cdf and LCR of this SIR process are given in \cite{Psomas23}. Substituting those expressions into \eqref{s*} yields
\begin{align}\label{SIRE}
F_{S^\star}^{\mathrm{SIR}_E}(s_{\text{th}})
\;\approx\;
F_{S(l)}^{\mathrm{SIR}_E}(s_{\text{th}})\,
\exp\!\left(
-\frac{L\,\mathrm{LCR}_{S(l)}^{\mathrm{SIR}_E}(s_{\text{th}})}{F_{S(l)}^{\mathrm{SIR}_E}(s_{\text{th}})}
\right),
\end{align}
where
\begin{align}
F_{S(l)}^{\mathrm{SIR}_E}(s_{\text{th}})
&= 1-\left(\frac{\Lambda}{\Lambda+s_{\text{th}}}\right)^{N},
\end{align}
and
\begin{align}\label{LCR_SIR_rayleighEqual}
\mathrm{LCR}_{S(l)}^{\mathrm{SIR}_E}(s_{\text{th}})
&= \frac{\Gamma\!\left(N+\tfrac{1}{2}\right)}{\Gamma(N)}
\sqrt{\frac{2b\, s_{\text{th}}}{\pi\Lambda}}
\left(\frac{\Lambda}{\Lambda+s_{\text{th}}}\right)^{N},
\end{align}
with $\Gamma(\cdot)$ the gamma function and $\Lambda$ the desired-to-interference power ratio (as defined in the main text).

\section{cdf of SINR for Equal-Power Interferers}\label{app:sinr_equal_power}
The superscript $\mathrm{SINR}_{\mathrm{E}}$ is used to denote the SINR for Rayleigh fading with $N$ \emph{equal-power} interferers. The cdf and LCR of this SINR process are given in \cite{Annavajjala_IMC_2010}. Substituting these into \eqref{s*} gives
\begin{align}\label{SINRME}
F_{S^\star}^{\mathrm{SINR}_E}(s_{\text{th}})
\;\approx\;
F_{S(l)}^{\mathrm{SINR}_E}(s_{\text{th}})\,
\exp\!\left(
-\frac{L\,\mathrm{LCR}_{S(l)}^{\mathrm{SINR}_E}(s_{\text{th}})}{F_{S(l)}^{\mathrm{SINR}_E}(s_{\text{th}})}
\right),
\end{align}
where
\begin{align}
F_{S(l)}^{\mathrm{SINR}_E}(s_{\text{th}})
&= 1-\exp\!\left(-\frac{s_{\text{th}}}{\gamma_0}\right)
\left(\frac{\Lambda}{\Lambda+s_{\text{th}}}\right)^N,
\end{align}
and
\begin{align}\label{lcr_equal}
\mathrm{LCR}_{S(l)}^{\mathrm{SINR}_E}(s_{\text{th}})
&= \sqrt{\frac{2b\, s_{\text{th}}}{\pi\gamma_0}}
\frac{\exp\!\left(W-\frac{s_{\text{th}}}{\gamma_0}\right)}{\Gamma(N)\sqrt{W}}
\left(\frac{\Lambda}{\Lambda+s_{\text{th}}}\right)^N\nonumber\\
&\quad\times \sum_{j=0}^{N-1}(-W)^{N-1-j}\binom{N-1}{j}
\Gamma\!\left(j+\tfrac{3}{2},W\right),
\end{align}
with $\gamma \triangleq \beta E_x/\sigma^2$, $\Lambda \triangleq \beta_0 E_{x_0}/(\beta E_x)$, and $W\triangleq 1/\gamma$. Here $\Gamma(\cdot,\cdot)$ denotes the upper incomplete gamma function.

\section{SIR Analysis under Ricean Fading}\label{Appendix:SIR_rician}
In this appendix we outline the derivation of the LCR used in \eqref{lcr_final_ric}. We work in the spatial domain (the antenna position $l$), and we use $\dot{(\cdot)}\triangleq\tfrac{d}{dl}(\cdot)$.

With a single interferer, the instantaneous SIR is
\begin{equation}
    S(l) = \frac{E_{x_{0}} |h_0(l)|^2}{E_{x_{1}}|h_1(l)|^2}.
\end{equation}
Let $r_0(l)=|h_{0}(l)|$, $r_1(l)=|h_{1}(l)|$, and define the ratio process $R(l)\triangleq r_0(l)/r_1(l)$. Then upcrossings of $S(l)$ across $s_{\text{th}}$ correspond to upcrossings of $R(l)$ across
\begin{align}
    r_{\text{th}}\;\triangleq\;\sqrt{\frac{s_{\text{th}}E_{x_{1}}}{E_{x_{0}}}}.
\end{align}

\subsection*{A. Local expansion of $r_0(l)$}
Let
\begin{align}
\alpha \triangleq \sqrt{\frac{\beta_0 K}{K+1}},\qquad
\gamma \triangleq \sqrt{\frac{\beta_0}{K+1}},
\end{align}
so that the Ricean desired channel in \eqref{ric_channel} can be written as
\begin{equation}
    h_{0}(l) = \alpha e^{-j\Phi l} + \gamma u(l),
\end{equation}
with $u(l)\sim\mathcal{CN}(0,1)$. For a small displacement $\tau$,
\begin{align}
    h_{0}(l+\tau)
    &= \alpha e^{-j\Phi(l+\tau)} + \gamma u(l+\tau)\nonumber\\
    &= \alpha e^{-j\Phi l}\big(1-j\Phi\tau+o(\tau)\big) + \gamma u(l+\tau).
\label{h0(l+tau)}
\end{align}
Moreover, using the standard decomposition
$u(l+\tau)=\rho(\tau)u(l)+\sqrt{1-|\rho(\tau)|^2}\,e$ with $e\sim\mathcal{CN}(0,1)$ and the small-$\tau$ expansion \eqref{correlation2}, we obtain
\begin{align}\label{u(l+tau)}
 u(l+\tau) = \big(1-b\tau^2\big)u(l) + \sqrt{2b}\,\tau\,e + o(\tau),
\end{align}
where the term $-b\tau^2u(l)$ is of higher order and will not contribute to the first-order derivative.
Substituting \eqref{u(l+tau)} into \eqref{h0(l+tau)} and retaining $O(\tau)$ terms gives
\begin{align}
    h_{0}(l+\tau)
    &= h_0(l) + \tau\Big(-j\alpha\Phi e^{-j\Phi l}+\gamma\sqrt{2b}\,e\Big)+o(\tau).
\end{align}
Using $|1+\tau z|=1+\tau\Re\{z\}+o(\tau)$, we obtain the first-order expansion of the envelope,
\begin{align}
|h_0(l+\tau)|
&= |h_0(l)|\Bigg[1+\tau\Re\!\left(\frac{-j\alpha\Phi e^{-j\Phi l}+\gamma\sqrt{2b}\,e}{h_0(l)}\right)\Bigg]\nonumber\\
&+ o(\tau).
\label{absh}
\end{align}
Writing $h_0(l)=|h_0(l)|e^{j\theta(l)}$ and neglecting $o(\tau)/\tau$ terms yields
\begin{align}\label{h(t+tau)2}
\dot{r}_0(l)
= \Re\!\Big( -j\alpha\Phi e^{-j(\Phi l+\theta(l))} + \gamma\sqrt{2b}\,e\,e^{-j\theta(l)}\Big).
\end{align}
Since $e\,e^{-j\theta(l)}\overset{d}{=}e$, we can write
\begin{align}
\dot{r}_0(l) \sim \mathcal{N}(\mu,\,\gamma^2 b),
\end{align}
where
\begin{align}
\mu \triangleq \Re\!\left\{-j\alpha\Phi e^{-j(\Phi l+\theta(l))}\right\}.
\end{align}

For the Rayleigh interferer, the envelope derivative satisfies
$\dot{r}_1(l)\sim\mathcal{N}(0,\,b\beta_1)$ \cite{Zlatanov_ICL_2008}.

\subsection*{B. LCR of the ratio process}
Since $R(l)=r_0(l)/r_1(l)$,
\begin{align}
\dot{R}(l)=\frac{\dot{r}_0(l)r_1(l)-\dot{r}_1(l)r_0(l)}{r_1^2(l)}.
\end{align}
Conditioned on $R(l)=r_{\text{th}}$ and $r_1(l)=y$, we have $r_0(l)=r_{\text{th}}y$ and
\begin{align}
\dot{R}(l)\;\sim\;\mathcal{N}\!\left(\frac{\mu}{y},\;\frac{\gamma^2 b + b\beta_1 r_{\text{th}}^2}{y^2}\right).
\end{align}
The upcrossing LCR of $R(l)$ across $r_{\text{th}}$ is
\begin{align}\label{rdotr}
  \mathrm{LCR}_{R(l)}(r_{\text{th}})
  = \int_0^{\infty} \dot{r}\, f_{R,\dot{R}}(r_{\text{th}},\dot{r})\, d\dot{r}.
\end{align}
To evaluate $f_{R,\dot{R}}$, we condition on $r_1(l)=y$ and the phase variable and then marginalize,
\begin{align}\label{frr}
    f_{\dot{R},R}(\dot{r},r_{\text{th}})
    &= \int_0^{2\pi} \int_0^{\infty} f_{\dot{R}|R,r_1,\theta}(\dot{r}|r_{\text{th}},y,\theta)\,\nonumber\\
    &\times f_{R,r_1,\theta}(r_{\text{th}},y,\theta)\, dy\, d\theta.
\end{align}
The conditional density is Gaussian with the parameters above, hence
\begin{align}\label{dotr}
    f_{\dot{R}|R,r_1,\theta}(\dot{r}|r_{\text{th}},y,\theta)
    = \frac{y}{\sqrt{\pi K_1}}
    \exp\!\left( -\frac{(\dot{r}-\mu/y)^2y^2}{K_1}\right),
\end{align}
where $K_1\triangleq 2\big(\gamma^2 b + b\beta_1 r_{\text{th}}^2\big)$.
Substituting \eqref{dotr} into \eqref{rdotr} and evaluating the integral over $\dot{r}$ using \cite[Eq.~(3.462-5)]{Gradshteyn_book_2007} yields
\begin{align}\label{int2}
    \int_0^{\infty} \dot{r}\,f_{\dot{R}|R,r_1,\theta}(\dot{r}|r_{\text{th}},y,\theta)\,d\dot{r}
    &= \frac{1}{2y}\Bigg(\sqrt{\frac{K_1}{\pi}}\,e^{-\mu^2/K_1}\nonumber\\
    &+\mu\bigg(1+\mathrm{erf}\!\left(\frac{\mu}{\sqrt{K_1}}\right)\bigg)\Bigg).
\end{align}
Next, conditioned on $r_1(l)=y$, the desired envelope/phase pair follows a scaled Ricean distribution, and the required joint density can be obtained from \cite{Kenneth_book_1974}. Combining with the Rayleigh density of $r_1(l)$ gives
\begin{align}\label{int3}
    f_{R,r_1,\theta}(r_{\text{th}},y,\theta)
    &= \frac{2y^3 r_{\text{th}}(K+1)}{\pi\beta_0\beta_1}\,e^{-K}\,e^{-K_2y^2}\,e^{K_3y},
\end{align}
where $K_2 \triangleq \frac{(K+1)r_{\text{th}}^2}{\beta_0}+\frac{1}{\beta_1}$ and
$K_3 \triangleq r_{\text{th}}\sqrt{\frac{K(K+1)}{\beta_0}}\cos(\theta-\Phi l)$.
Substituting \eqref{int2} and \eqref{int3} into \eqref{frr} and \eqref{rdotr} yields \eqref{lcr_ric}, shown at the top of the next page.
Finally, integrating over $y$ using \cite[Eq.~(3.462-5) and Eq.~(3.462-7)]{Gradshteyn_book_2007} and converting from $R(l)$ back to $S(l)$ gives \eqref{lcr_final_ric}. Since $S(l)$ follows a noncentral $F$-distribution, its cdf is available in \cite[Chap.~30]{JK2}, which leads to \eqref{cdf_sir_ric}.

\begin{figure*}[t]
\normalsize
\begin{align}\label{lcr_ric}
    \mathrm{LCR}_{R(l)}(r_{\text{th}})
    = \int_0^{2\pi}\!\int_0^{\infty}
    \Bigg(\frac{1}{y}\Big(\tfrac{1}{2}\sqrt{\tfrac{K_1}{\pi}}\,e^{-\mu^2/K_1} + \tfrac{\mu}{2}\big(1+\mathrm{erf}(\tfrac{\mu}{\sqrt{K_1}})\big)\Big)
    \frac{2y^3 r_{\text{th}}(K+1)}{\pi\beta_0\beta_1} e^{-K}e^{-K_2y^2}e^{K_3y}\Bigg)\,dy\,d\theta.
\end{align}
\hrulefill
\end{figure*}

\section{Approximation of the cdf of SNR for a Single Fluid Antenna with a Fixed Antenna}\label{appendix:cdf_onefixed_onefluid}
We first consider the unequal-power case $\beta_0\neq\beta_f$.
Let $G(l)\triangleq |h_f|^2+|h_0(l)|^2$ so that $S(l)=\tfrac{E_{x_0}}{\sigma^2}G(l)$. Then the LCR of $S(l)$ across $s_{\text{th}}$ equals the LCR of $G(l)$ across
$g_{\text{th}}\triangleq \tfrac{\sigma^2 s_{\text{th}}}{E_{x_0}}$.
The LCR of $G(l)$ is
\begin{align}
\mathrm{LCR}_{G(l)}(g_{\text{th}}) = \int_0^{\infty} \dot{g}\, f_{G,\dot{G}}(g_{\text{th}},\dot{g})\,d\dot{g}.
\end{align}
Let $r(l)=|h_0(l)|$. Since $G(l)=|h_f|^2+r^2(l)$, we have $\dot{G}(l)=2r(l)\dot{r}(l)$.
Under Rayleigh fading with spatial correlation satisfying \eqref{correlation2}, $\dot{r}(l)\sim\mathcal{N}(0,b\beta_0)$, hence
$\dot{G}(l)\,\big|\,r\sim\mathcal{N}(0,4r^2 b\beta_0)$.
Therefore,
\begin{align}
\mathrm{LCR}_{G(l)}(g_{\text{th}})
&=\int_0^{\infty}\!\int_0^{\infty} \dot{g}\, f_{\dot{G}|G,r}(\dot{g}|g_{\text{th}},r)\, f_{G,r}(g_{\text{th}},r)\,dr\,d\dot{g}.
\end{align}
Using
\begin{align}
    f_{\dot{G}|G,r}(\dot{g}|g_{\text{th}},r)
    = \frac{1}{\sqrt{8\pi r^2 b\beta_0}}\exp\!\left(-\frac{\dot{g}^2}{8r^2 b\beta_0}\right)
\end{align}
and \cite[Eq.~(3.326-2)]{Gradshteyn_book_2007}, we obtain
\begin{align}\label{lcrg}
    \mathrm{LCR}_{G(l)}(g_{\text{th}})
    = \int_0^{\infty} r\sqrt{\frac{2b\beta_0}{\pi}}\, f_{G,r}(g_{\text{th}},r)\,dr.
\end{align}
To compute $f_{G,r}$, treat $(G,r)$ as a transformation of $(X,Y)=(|h_f|^2,r^2)$.
Standard transformation theory \cite{papoulis} gives
\begin{align}\label{fgr}
    f_{G,r}(g_{\text{th}},r)
    = \frac{2r}{\beta_f\beta_0}
    \exp\!\left(-\frac{g_{\text{th}}}{\beta_f}\right)
    \exp\!\left(-\varsigma r^2\right),
\end{align}
where $\varsigma\triangleq \tfrac{1}{\beta_0}-\tfrac{1}{\beta_f}$ and $0\le r\le \sqrt{g_{\text{th}}}$.
Substituting \eqref{fgr} into \eqref{lcrg} yields
\begin{align}\label{lcrg2}
\mathrm{LCR}_{G(l)}(g_{\text{th}})
&= 2\sqrt{\frac{2b\beta_0}{\pi}}\,\frac{e^{-g_{\text{th}}/\beta_f}}{\beta_f\beta_0}
\int_0^{\sqrt{g_{\text{th}}}} r^2 e^{-\varsigma r^2}\,dr.
\end{align}
Evaluating the integral using \cite[Eq.~(3.326-2)]{Gradshteyn_book_2007} gives the result in \eqref{lcr1fix} after converting back via
$\mathrm{LCR}_{S(l)}^{\mathrm{FF,uneq}}(s_{\text{th}})=\mathrm{LCR}_{G(l)}\!\big(\tfrac{\sigma^2 s_{\text{th}}}{E_{x_0}}\big)$.

For the marginal cdf,
\begin{align}
F_{S(l)}^{\mathrm{FF,uneq}}(s_{\text{th}})
&= \mathbb{P}\!\left(G(l) < \frac{\sigma^2 s_{\text{th}}}{E_{x_0}}\right).
\end{align}
Since $G(l)$ is the sum of two independent exponentials, it is hypoexponential \cite[Chap.~28]{hypo}; using \cite{Norman_book_1971} yields \eqref{cdf_fixed_fluid}.

\paragraph{Equal powers ($\beta_f=\beta_0=\beta$)}
When $\varsigma=0$, the integral in \eqref{lcrg2} reduces to $\int_0^{\sqrt{g_{\text{th}}}} r^2dr=g_{\text{th}}^{3/2}/3$, which yields \eqref{lcrfix1same}.
Moreover, $G(l)$ becomes Erlang (Gamma) with shape~2, so $S(l)$ in \eqref{snr_fixed} follows a scaled chi-square distribution with four degrees of freedom.
Combining the resulting marginal cdf with \eqref{s*} gives \eqref{FF_EqP}.

\section{LCR derivation of the SNR for a correlated moving antenna array}\label{Appendxi:LCR_array_de}
Consider a two-element rigid array that is translated along a track. Let
$\mathbf{h}(l)=[h_1(l),\,h_2(l)]^T$, where $h_i(l)$ is the channel to the $i$th array element at location $l$.
For notational simplicity we assume $h_i(l)\sim\mathcal{CN}(0,1)$ and reintroduce the average power scaling $\beta$ at the end.
Assume the small-distance correlation model \eqref{correlation2} and denote
\begin{align}
J\triangleq \mathbb{E}[h_1(l)h_2^*(l)] = J_0(2\pi\Delta).
\end{align}

\subsection*{A. Local correlation expansion}
The correlation between $h_1(l)$ and $h_2(l+\tau)$ is
\begin{align}
\mathbb{E}[h_1(l)h_2^*(l+\tau)]
&= \rho\!\left(\sqrt{\Delta^2+\tau^2}\right)\nonumber\\
&= J_0\!\left(2\pi\Delta+\frac{\pi\tau^2}{\Delta}\right)+o(\tau^2),
\end{align}
where we used $\sqrt{\Delta^2+\tau^2}=\Delta+\tfrac{\tau^2}{2\Delta}+o(\tau^2)$.
Using $\tfrac{d}{dx}J_0(x)=-J_1(x)$ \cite[Eq.~9.1.28]{AS}, we obtain
\begin{align}
J_0\!\left(2\pi\Delta+\frac{\pi\tau^2}{\Delta}\right)
= J_0(2\pi\Delta) - \frac{\pi}{\Delta}J_1(2\pi\Delta)\,\tau^2 + o(\tau^2).
\end{align}
Define
\begin{align}
 c_1 \triangleq \frac{\pi}{\Delta}J_1(2\pi\Delta),
\end{align}
so that $\rho(\sqrt{\Delta^2+\tau^2})=J-c_1\tau^2+o(\tau^2)$.

Let $\mathbf{v}(l)\triangleq [h_1(l),\,h_2(l),\,h_1(l+\tau),\,h_2(l+\tau)]^T$.
Up to $o(\tau^2)$, its covariance matrix can be written as
\begin{align}
\mathbf{R}
= \begin{bmatrix}
\mathbf{\Sigma} & \mathbf{\Sigma}-\tau^2\mathbf{A}\\
\mathbf{\Sigma}-\tau^2\mathbf{A} & \mathbf{\Sigma}
\end{bmatrix} + o(\tau^2),
\end{align}
where $\mathbf{\Sigma}=\begin{bmatrix}1 & J\\ J & 1\end{bmatrix}$ and
$\mathbf{A}=\begin{bmatrix}b & c_1\\ c_1 & b\end{bmatrix}$.
A block Cholesky factorization of the $O(\tau^2)$ approximation yields
\begin{align}
\begin{bmatrix} \mathbf{h}(l) \\ \mathbf{h}(l+\tau)\end{bmatrix}
= \begin{bmatrix} \mathbf{F} & \mathbf{0} \\ \mathbf{G} & \mathbf{H}\end{bmatrix}
\begin{bmatrix} \mathbf{u}_1 \\ \mathbf{u}_2\end{bmatrix},
\end{align}
where $\mathbf{u}_1,\mathbf{u}_2\sim\mathcal{CN}(\mathbf{0},\mathbf{I}_2)$ are independent and
\begin{align}
\mathbf{F} &= \begin{bmatrix} 1 & 0 \\ J & \sqrt{1-J^2} \end{bmatrix}, \\
\mathbf{G} &= \big(\mathbf{F}^{-1}\mathbf{\Sigma} - \tau^2 \mathbf{F}^{-1}\mathbf{A}\big)^{H}, \\
\mathbf{H} &= \tau \mathbf{C}.
\end{align}
with
\begin{align}
\mathbf{C}=\sqrt{2}\begin{bmatrix}
\sqrt{b} & 0\\
\frac{c_1}{\sqrt{b}} & \sqrt{\frac{b^2-c_1^2}{b}}
\end{bmatrix}.
\end{align}
Substituting and simplifying gives the convenient recursion
\begin{align}\label{hrec}
\mathbf{h}(l+\tau) = (\mathbf{I}-\tau^2\mathbf{B})\mathbf{h}(l) + \tau\mathbf{C}\mathbf{u}_2,
\end{align}
where $\mathbf{B}=\mathbf{A}^H\mathbf{\Sigma}^{-1}$.

\subsection*{B. Derivative of the post-combining SNR}
For general average power $\beta$, the MRC SNR is
$S(l)=\tfrac{\beta E_{x_0}}{\sigma^2}\,\mathbf{h}^H(l)\mathbf{h}(l)$.
Using \eqref{hrec} and keeping the leading $O(\tau)$ term yields
\begin{align}
\dot{S}(l)
&= \lim_{\tau\to 0} \frac{\beta E_{x_0}}{\sigma^2}\,\frac{\mathbf{h}^H(l+\tau)\mathbf{h}(l+\tau)-\mathbf{h}^H(l)\mathbf{h}(l)}{\tau}\nonumber\\
&= \frac{\beta E_{x_0}}{\sigma^2}\Big(\mathbf{h}^H(l)\mathbf{C}\mathbf{u}_2 + \mathbf{u}_2^H\mathbf{C}^H\mathbf{h}(l)\Big).
\end{align}
Conditioned on $\mathbf{h}(l)=\mathbf{h}$, the complex variable $\mathbf{h}^H\mathbf{C}\mathbf{u}_2\sim\mathcal{CN}(0,\mathbf{h}^H\mathbf{C}\mathbf{C}^H\mathbf{h})$, hence
\begin{align}\label{snr_dot_array2-simp}
\dot{S}(l)\,\big|\,\mathbf{h}
\sim \mathcal{N}\!\left(0,\;\left(\frac{\beta E_{x_0}}{\sigma^2}\right)^{\!2}\,\mathbf{h}^H\mathbf{Q}\mathbf{h}\right),
\end{align}
where $\mathbf{Q}\triangleq 2\mathbf{C}\mathbf{C}^H = 4\mathbf{A}$.

\subsection*{C. Quadratic-form representation}
Write $\mathbf{h}(l)=\mathbf{F}\mathbf{u}$ with $\mathbf{u}\sim\mathcal{CN}(\mathbf{0},\mathbf{I}_2)$.
Then
\begin{align}\label{qf1}
\mathbf{h}^H\mathbf{h} = \mathbf{u}^H\mathbf{\Sigma}\mathbf{u},
\end{align}
and
\begin{align}\label{qf2}
\mathbf{h}^H\mathbf{Q}\mathbf{h} = 4\,\mathbf{u}^H\mathbf{\Sigma}^{1/2}\mathbf{A}\mathbf{\Sigma}^{1/2}\mathbf{u}.
\end{align}
Both are Hermitian quadratic forms. Let $\lambda_1,\lambda_2$ denote the eigenvalues of $\mathbf{\Sigma}$, which are
\begin{align}
\lambda_1=1-J,\qquad \lambda_2=1+J.
\end{align}
Moreover, $\mathbf{\Sigma}^{1/2}\mathbf{A}\mathbf{\Sigma}^{1/2}$ shares eigenvectors with $\mathbf{\Sigma}$, so its eigenvalues can be written as
\begin{align}
\theta_1=(1-J)\Big(1-\tfrac{c_1}{b}\Big),\qquad
\theta_2=(1+J)\Big(1+\tfrac{c_1}{b}\Big).
\end{align}
Hence we can express
\begin{align}
S(l) &= c_0\,X,\qquad X\triangleq \lambda_1|u_1|^2+\lambda_2|u_2|^2,\nonumber\\
\dot{S}(l)\,\big|\,Y &= \mathcal{N}(0,\,a_1^2 Y),\qquad Y\triangleq \theta_1|u_1|^2+\theta_2|u_2|^2,
\end{align}
where $c_0\triangleq \beta E_{x_0}/\sigma^2$ and $a_1\triangleq 2\sqrt{b}\,\beta E_{x_0}/\sigma^2$.

\subsection*{D. LCR evaluation}
Using the standard LCR identity for a differentiable stationary process,
\begin{align}
\mathrm{LCR}_{S(l)}(s_{\text{th}})
&= \int_0^{\infty} \dot{s}\, f_{S,\dot{S}}(s_{\text{th}},\dot{s})\,d\dot{s}\nonumber\\
&= \int_{-\infty}^{\infty}\! \frac{a_1\sqrt{y}}{\sqrt{2\pi}}\, f_{S,Y}(s_{\text{th}},y)\,dy,
\label{eq:lcr_SY}
\end{align}
where we used \cite[Eq.~(3.326-2)]{Gradshteyn_book_2007} for the inner Gaussian integral.
Since $S=c_0X$, we have
\begin{align}
 f_{S,Y}(s_{\text{th}},y)=\frac{1}{c_0} f_{X,Y}(s_{\text{th}}/c_0, y).
\end{align}
The joint density of $(X,Y)$ follows from a linear transformation of $(|u_1|^2,|u_2|^2)$.
Using transformation theory \cite{papoulis}, one obtains
\begin{align}\label{eq:fXY}
 f_{X,Y}(x,y)=\frac{1}{D}\exp\!\left(-\frac{(\theta_2-\theta_1)x}{D}\right)
 \exp\!\left(-\frac{(\lambda_2-\lambda_1)y}{D}\right),
\end{align}
valid over $0<\tfrac{\theta_1}{\lambda_1}x<y<\tfrac{\theta_2}{\lambda_2}x$, where
$D\triangleq \lambda_1\theta_2-\lambda_2\theta_1$.
Substituting \eqref{eq:fXY} into \eqref{eq:lcr_SY} and carrying out the remaining integral yields the closed-form LCR expression in \eqref{corr_array_lcr}.

\section*{Acknowledgment}
The authors thank Prof. Malin Premaratne for valuable comments and suggestions, particularly in the derivation of the cdf approximation for fluid antenna systems.
\bibliographystyle{IEEEtran}
\bibliography{references}

\end{document}